\documentclass[10pt]{article}

\usepackage[T1]{fontenc}
\usepackage[utf8]{inputenc}
\usepackage{newtxtext,newtxmath}   
\usepackage[letterpaper,left=1.5in,right=1.5in,top=1in,bottom=1in]{geometry}
\usepackage{setspace}
\usepackage{titlesec}
\usepackage{hyperref}
\usepackage{xcolor}
\usepackage{longtable}
\usepackage{booktabs}
\usepackage{array}
\usepackage{fancyhdr}
\usepackage{graphicx}
\usepackage{enumitem}
\usepackage[hang,flushmargin]{footmisc}
\usepackage{tocloft}
\usepackage{indentfirst}

\hypersetup{
  colorlinks=true,
  linkcolor=black,
  citecolor=black,
  urlcolor=blue,
}

\titleformat{\section}
  {\centering\normalfont\normalsize\scshape}
  {\Roman{section}.}{0.5em}{\MakeUppercase}
\titlespacing{\section}{0pt}{24pt}{12pt}

\titleformat{\subsection}
  {\normalfont\normalsize\itshape}
  {\Alph{subsection}.}{0.5em}{}
\titlespacing{\subsection}{0pt}{18pt}{6pt}

\titleformat{\subsubsection}
  {\normalfont\normalsize\itshape}
  {\arabic{subsubsection}.}{0.5em}{}
\titlespacing{\subsubsection}{0pt}{12pt}{6pt}

\renewcommand{\footnotesize}{\small}

\newcommand{\ia}{\textit{i}}
\newcommand{\ib}{\textit{ii}}
\newcommand{\ic}{\textit{iii}}

\begin{document}

\thispagestyle{empty}
\begin{center}
  {\large \textbf{Digital Speech Acts Retain Control of Copyright\\ with People, Not Platforms}}\\[12pt]
  By James Golike and Ehud Shapiro\textsuperscript{*}\\[24pt]
\end{center}

\begin{abstract}
Legal precedents protect computer code as copyrightable expression. They have enabled centralized digital platforms---operating from corporate servers that hold all user data---to construct private governance regimes through the interaction of copyright, contract, and technical architecture: people who create virtually all platform value must surrender effective copyright control through Terms of Service agreements as a condition of participation.

In contrast, grassroots platforms consist of cryptographically-identified people operating their networked smartphones independently of any server or global resource; each person holds their own data on their own device, with no third party in possession or intermediation. Here, we define the notion of a \textit{digital speech act}---a deliberate volitional act by a person of cryptographically signing personal content with the person's private key, carried out on the person's own device---through which the person simultaneously establishes attribution, accountability, and authorship over the signed content. We contend that (\ia) digital speech acts qualify for copyright protection under existing U.S.\ precedent: \textit{Burrow-Giles} locates authorship in volitional creative choices despite mechanical or algorithmic processes, \textit{Feist} supplies the minimal-creativity threshold, and persistent device storage satisfies the Copyright Act's fixation requirement; (\ib) the digital social contract underlying grassroots platforms preserves this copyright by design—signed content cannot be unbundled from its signature, and the full provenance chain accumulates as content is forwarded—so that copyright ownership and physical possession of authenticated digital expressions coalesce in the person; and (\ic) this coalescence of legal ownership and physical possession provides the foundations for digital sovereignty and democratic self-governance.
\end{abstract}

\vfill

\noindent\rule{2in}{0.4pt}

\smallskip

{\footnotesize * Ehud Shapiro is at the London School of Economics and the Weizmann Institute of Science.
The authors used a generative AI tool (Claude by Anthropic) for drafting, restructuring, research and source identification, copy-editing, citation formatting, and \LaTeX{} formatting during the preparation of this manuscript.}

\newpage
\setcounter{page}{1}

\begin{center}
\textsc{Table of Contents}
\end{center}
\smallskip
\makeatletter
\@starttoc{toc}
\makeatother

\newpage


\section{Introduction}\label{sec:intro}

The question of whether cryptographically authenticated digital expression qualifies as a copyrightable work under existing United States copyright law has acquired increasing practical significance as digital communication becomes authenticated through cryptographic identity rather than centralized intermediaries. Yet copyright doctrine has never directly addressed whether expression fixed, authenticated, and attributed through cryptographic means constitutes a copyrightable work. The significance of this doctrinal question extends well beyond copyright doctrine itself. Contemporary digital platforms already recognize users as the copyright owners of their expressive works, yet routinely require those same users to grant extraordinarily broad licenses through Terms of Service agreements as a condition of participation, placing copyright and contract at the center of platform governance.

Platforms wield two legal instruments in tandem: copyright to protect their code, and contract to extract user content. The legal framework enabling this asymmetry emerged through case-by-case litigation over individual software disputes in the 1980s and 1990s—decisions that prioritized developer interests over user rights, validated adhesion contracts on economic efficiency grounds, and granted comprehensive protection to all forms of computer code.

Consider what happens when a person posts on Facebook, uploads to YouTube, or shares on X. Copyright law recognizes the person as the author of their creation, granting exclusive rights under 17 U.S.C. § 106 to reproduce, distribute, and display the work. But to participate in digital life—to connect with friends, share with family, engage in public discourse—a person must first agree to Terms of Service granting platforms "worldwide, non-exclusive, transferable, sub-licensable and royalty-free" licenses to everything they create.\footnote{See Meta, Terms of Service (effective Mar. 4, 2026), https://www.facebook.com/terms/ (last visited May 18, 2026); YouTube, Terms of Service (effective Jan. 1, 2025), https://www.youtube.com/static?template=terms (last visited May 18, 2026); X, Terms of Service (effective Nov. 24, 2024), https://x.com/en/tos (last visited May 18, 2026).} In return, the person receives revocable access to platform service—a license the platform can terminate at any time, for any reason, often without explanation.

Meanwhile, platforms assert the full force of copyright law to protect their own code. Every API, every algorithm, every line of machine-readable object code receives comprehensive copyright protection as "literary works." The asymmetry is stark: platforms use copyright defensively over their code and contract offensively over user content. Platform code is protected property; user content is extracted capital.

This was not a technological inevitability. The architecture of platform governance emerged through specific historical and technical conditions during the Internet's formative years. The 1976 Copyright Act extended copyright protection to computer programs while leaving critical questions unresolved—whether code in machine-readable form was copyrightable, whether programs stored in RAM were "fixed," and where to draw the line between copyrightable expression and uncopyrightable ideas.\footnote{See infra Part II.B.} Congress delegated these questions to the courts.\footnote{\textit{Id.}}

Federal courts frequently resolved these questions in favor of software developers. \textit{Apple Computer v. Franklin Computer} (1983) held that all computer code—even purely functional operating systems in machine-readable object code—qualifies for copyright protection.\footnote{\textit{Apple Computer, Inc. v. Franklin Computer Corp.}, 714 F.2d 1240, 1249 (3d Cir. 1983).} \textit{MAI Systems v. Peak Computer} (1993) held that even temporary loading of software into RAM creates a "fixed" copy subject to copyright, and that software users are "licensees" bound by all license terms, not "owners" entitled to copyright exemptions.\footnote{\label{fn:mai}
\textit{MAI Systems Corp. v. Peak Computer, Inc.},
991 F.2d 511, 518 (9th Cir. 1993).} \textit{ProCD v. Zeidenberg} (1996) held shrinkwrap licenses enforceable under the UCC, reasoning that "competition among vendors, not judicial revision of a package's contents, is how consumers are protected in a market economy."\footnote{\label{fn:procd}
ProCD, Inc. v. Zeidenberg, 86 F.3d 1447, 1451 (7th Cir. 1996); see infra Part II.C.} These decisions prioritized developer economic interests over user rights.\footnote{See infra Part II.}

This legal framework developed alongside the Internet's privatization and commercialization in the late 1980s and early 1990s, within a permissive U.S. regulatory environment that treated Internet service providers as "value-added services" requiring minimal state intervention.\footnote{\label{fn:radu}Roxana Radu, \textit{Negotiating Internet Governance} 61 (2019).} Not all courts followed this approach—subsequent decisions imposed notice and assent requirements that limited enforceability—but platforms adapted their Terms of Service structures to satisfy these standards while preserving the asymmetric license exchange.\footnote{See infra Part II.C (discussing \textit{Specht v. Netscape}, \textit{Klocek v. Gateway}, \textit{Bragg v. Linden Research}, and platform adaptation through hybridwrap mechanisms).}

These decisions established what we term a "copyright paradox"\footnote{\label{fn:prehistory}
James Golike, \textit{A Pre-History of Platform Power: A Critical Examination
of Early U.S. Legal Discourse Defining Global Platform-User Contractual
Relations} (2025) (LSE MSc dissertation) (on file with author).}: software licensing simultaneously evades and depends upon copyright law, creating private governance structures that exceed the intended scope of intellectual property law.\footnote{See infra Part~\ref{sec:platform-regime}.} Platforms evade copyright's limitations by characterizing transactions as licenses rather than sales, denying users the exemptions the Copyright Act grants to owners. Yet they depend on copyright's enforcement power to validate these licenses as enforceable contracts. The result is a private governance regime through which platforms obtain quasi-sovereign control over essential digital communication infrastructures—a characterization advanced by scholars across antitrust law, information law, and critical platform studies.\footnote{See, e.g., \label{fn:khan-antitrust}Lina M. Khan, \textit{Amazon's Antitrust Paradox}, 126 Yale L.J. 710 (2017) (analyzing platform monopoly power over essential infrastructure); \label{fn:balkin-fiduciaries}Jack M. Balkin, \textit{Information Fiduciaries and the First Amendment}, 49 U.C. Davis L. Rev. 1183 (2016) (platforms as exercising quasi-governmental functions); \label{fn:cohen-truth}Julie E. Cohen, \textsc{Between Truth and Power} 37--65 (2019) (platforms as infrastructural power); \label{fn:pasquale-blackbox}Frank Pasquale, \textsc{The Black Box Society} 19--58 (2015) (platform control resembling sovereign authority).}

This legal framework, built upon decisions addressing individual software disputes in a pre-platform world, has scaled to become the architecture which governs billions globally through "hybridwrap" Terms of Service agreements.\footnote{See infra Part~\ref{sec:platform-regime}.E.} Every major platform deploys the same structure: comprehensive copyright protection for platform code, mandatory content licenses from users, and revocable access as the only consideration for use.

Global regulatory responses, such as the EU's GDPR and Digital Services Act, have sought to impose protective rules on digital users while leaving the fundamental copyright appropriation structure intact. Platforms are permitted to accumulate concentrated control of user-generated content and data, with regulators then attempting to impose \textit{ex post} protections—privacy rules, content moderation requirements, data portability rights, and data localization requirements. Yet these regulatory interventions cannot address the underlying dilemma: platforms retain comprehensive content licenses and physical possession of user data regardless of protective rules or territorial mandates. Regulatory protections address symptoms rather than causes. The cause is the legal and technical architecture which enables copyright appropriation and the separation of ownership from possession. Users may nominally own copyright in their content and data, while platforms receive broad licenses to said content and control the data through centralized servers.

Grassroots networks provide a contrasting architectural arrangement. These serverless, distributed, permissionless systems operate solely on the networked smartphones of cryptographically-identified participants, and have been formally specified as multiagent systems of volitional transactions.\footnote{\label{note:vmat}Andy Lewis-Pye \& Ehud Shapiro, \textit{Volitional Multiagent Atomic Transactions: Describing People and their Machines} (2026) (arXiv preprint arXiv:2604.25596).} In such systems, each person remains in possession of their own data, with no third party in possession or intermediation.\footnote{\label{note:gsn}Ehud Shapiro, \textit{Grassroots Social Networking: Where People Have Agency over Their Personal Information and Social Graph} (2024) (arXiv preprint arXiv:2306.13941).}
 Unlike centralized platforms (Facebook, X) or federated systems (Mastodon), grassroots platforms have no external legal entity that can draft Terms of Service or extract comprehensive content licenses from users as a condition of access. Each person stores their data on their own smartphone, with no third party able to obtain access unless explicitly granted. The absence of an operating entity has a direct legal consequence: there is no legal person positioned to require users to grant broad copyright licenses through contractual frameworks as a prerequisite for participation.\footnote{See infra Part~\ref{sec:grassroots}.}

Within this architectural setting, the central object is the \textbf{digital speech act}—a deliberate volitional act by a person, performative in the sense of Searle and Austin,\footnote{See infra Part IV.D.1.} and realized by cryptographically signing personal content with the person's private key. Through this act the person simultaneously establishes attribution (this content is verifiably mine), accountability (I stand behind it and can be held responsible for it), and authorship (I own the resulting expression).

This Article makes the case that digital speech acts should qualify for copyright protection under existing U.S. precedent. Not through the platform-era authorities that enabled extraction, but through foundational copyright principles predating platforms by decades or centuries. The primary precedent is \textit{Burrow-Giles Lithographic Co. v. Sarony} (1884), which remains significant not merely because it involved photography, but because it illustrates how copyright law confronts novel technologies. 

Faced with a medium that mechanically captured images, the Supreme Court did not focus on the camera's operation as such. Instead, it asked whether the expressive elements embodied in the resulting work originated with a human author. The Court concluded that authorship existed because the photographer selected the subject, pose, costume, lighting, and arrangement, while the camera merely executed those choices mechanically.\footnote{\label{fn:burrow-giles}
\textit{Burrow-Giles Lithographic Co. v. Sarony},
111 U.S. 53, 60 (1884); see infra Part~\ref{sec:core}.}

Digital speech acts present a similar doctrinal question. The relevant inquiry is not whether cryptographic systems participate in the creation of an expression, but whether the expressive choices embodied in the digital speech act originate with the human participant or with the technology itself. A person creating a digital speech act chooses what content to sign, when to sign it, what communicative act to perform, and which cryptographic identity to use. Cryptographic algorithms execute those choices mathematically—just as cameras execute photographers' choices mechanically—but make no creative decisions themselves. 

Once authorship is established, \textit{Feist Publications, Inc. v. Rural Telephone Service Co.} (1991) supplies the governing originality standard. \textit{Feist} reaffirmed that copyright requires only a minimal degree of creativity—an ``extremely low'' threshold satisfied by modest choices in selection, coordination, and arrangement.\footnote{\label{fn:feist}
\textit{Feist Publications, Inc. v. Rural Telephone Service Co.},
499 U.S. 340, 345 (1991); see infra Part~\ref{sec:core}.} The Copyright Act's fixation requirement likewise applies straightforwardly to digital speech acts stored persistently on participant devices.\footnote{17 U.S.C. \S~102(a).}

Taken together, these authorities demonstrate that digital speech acts satisfy the core requirements of copyrightability. Burrow-Giles locates authorship in the human expressive choices embodied in the work, Feist confirms that those choices readily exceed copyright's minimal originality threshold, and the Copyright Act's fixation requirement is satisfied through persistent device storage. Digital speech acts should therefore qualify for copyright protection under established copyright doctrine.

Copyright ownership alone, however, is insufficient without architectural enforcement. The digital social contract underlying grassroots networks—a voluntary agreement among people, specified, fulfilled, and enforced through code\footnote{\label{note:cardelli}Luca Cardelli et al., \textit{Digital Social Contracts: A Foundation for an Egalitarian and Just Digital Society} (2020) (arXiv preprint arXiv:2005.06261).}—specifies how digital speech acts are handled. It enforces two rules as machine transactions guarded by the presence of a digital speech act: (1) \textbf{non-unbundling}—signed content cannot be separated from the signature when forwarded, ensuring the creator's copyright claim travels with the content; and (2) \textbf{provenance preservation}—the full chain of forwarders must be preserved, identifying everyone who forwarded the content.\footnote{See infra Part~\ref{sec:dsa}.} Together, these rules maintain the original copyright and establish clear accountability for both creators and forwarders. 

The result is the \textit{coalescence of ownership and possession}---a fundamental architectural difference from platform regimes. In grassroots architecture the person who owns the copyright in a digital speech act also holds the copyrighted expression on their own device. No third party possesses the content, and no Terms of Service can extract comprehensive licenses from what no third party possesses. Copyright ownership thus becomes practically enforceable---not merely a nominal right surrendered through adhesive licensing, but actual control over reproduction, distribution, and use of one's own digital expressions.

The significance of recognizing digital speech acts as copyrightable expression extends beyond copyright doctrine itself. Copyright provides a legal basis for personal control over the creation, circulation, and attribution of expression. When persons both own and possess their digital expressions, participation in digital life no longer depends upon third-party custody, platform access, or the surrender of rights through contractual intermediation. 

The resulting independence suggests a broader conception of \textit{digital sovereignty}---the ability to conduct one's digital life free of third-party control, surveillance, manipulation, and rent-seeking.\footnote{\label{fn:shapiro-grassroots-systems}Ehud Shapiro, \textit{Grassroots Systems: Concept, Examples, Implementation and Applications}, in Proceedings of the 37th International Symposium on Distributed Computing (DISC 2023) 47:1--47:18 (2023) (also at arXiv:2301.04391).} Recognizing copyright in digital speech acts therefore carries broader economic and democratic implications, because it establishes a legal foundation upon which alternative forms of digital organization may be constructed. Rather than treating digital communication primarily as a source of value extraction mediated by platforms, it creates the possibility of organizing digital relationships around the rights and responsibilities of the persons who create and communicate digital expression. 

Whether such sovereignty ultimately requires forms of property ownership is a question with deep roots in democratic theory. Drawing on traditions associated with Locke,\footnote{\label{fn:locke}John Locke, \textsc{Second Treatise of Government} §§ 25--51 (C.B. Macpherson ed., 1980) (1690).} Rawls,\footnote{\label{fn:rawls-jf}John Rawls, \textsc{Justice as Fairness: A Restatement} 135--40 (Erin Kelly ed., 2001).} and Singer,\footnote{\label{fn:singer-entitlement}Joseph William Singer, \textsc{Entitlement: The Paradoxes of Property} (2000).} Part~\ref{sec:democratic-foundations} examines the relationship between property, sovereignty, and democratic participation in digital environments, while engaging substantial criticisms of ownership-based approaches from intellectual property and commons scholarship.

The argument that digital speech acts qualify for copyright protection under existing doctrine, however, faces serious objections. Critics may contend that digital speech acts are machine-generated rather than authored, functional rather than expressive, mechanically produced rather than creative, or too thinly original to warrant meaningful protection. Existing doctrine provides support for each of these concerns. Recent algorithmic generation cases (\textit{Thaler v. Perlmutter}, \textit{Naruto v. Slater}) deny copyright to machine-created outputs. Idea/expression doctrine (\textit{Baker v. Selden}) excludes methods from protection. Merger doctrine prevents protection when expression is input-dictated. Mechanical reproduction precedents (\textit{Bridgeman Art Library v. Corel Corp.}, \textit{Meshwerks, Inc. v. Toyota Motor Sales U.S.A., Inc.}) deny copyright to slavish copies lacking creativity. Thin copyright scholarship warns against minimal creativity enabling strategic overclaiming. Contract preemption (§301), first sale (§109), and copyright skeptics raise doctrinal and policy concerns. Part~\ref{sec:challenges} addresses each of these objections, testing the digital speech act framework against the principal doctrinal and policy challenges raised by existing copyright law.

The Article proceeds as follows. Part~\ref{sec:platform-regime} examines the current regime of platform copyright dominance, tracing how 1980s--90s copyright and software licensing precedents have scaled into mechanisms for global private governance. Part~\ref{sec:grassroots} describes grassroots platforms and the architectural features that preclude third-party possession and intermediation. Part~\ref{sec:dsa} defines digital speech acts theoretically and technically, situating them within both Searle and Austin's speech act theory and VMAT's volitional-transaction framework. Part~\ref{sec:core} develops the doctrinal argument, identifying \textit{Burrow-Giles} as the primary precedent for human authorship despite mechanical execution and drawing additional support from \textit{Feist}'s originality standard and the Copyright Act's fixation requirement. Part~\ref{sec:user-person} develops the distinction between legal persons and platform users, examining how contemporary platforms transform persons into contractually-defined users through Terms of Service while grassroots architectures reestablish authorship, ownership, and accountability through cryptographic identity and binding. Part~\ref{sec:ownership-sovereignty} develops the relationship between ownership and sovereignty in grassroots digital environments, examining how copyright, possession, and contractual governance combine to support independent participation outside intermediary-controlled systems. Part~\ref{sec:provenance} examines provenance preservation and how it maintains copyright while establishing accountability. Part~\ref{sec:challenges} addresses defensive strategies against anticipated legal challenges—algorithmic generation, idea/expression, mechanical reproduction, merger, thin copyright, contract preemption, first sale, and scholarly critics. Part~\ref{sec:democratic-foundations} develops the democratic foundations of distributed ownership, examining digital expression through theories of property, productive assets, democratic independence, and anti-domination. Part~\ref{sec:economic-implications} examines the economic implications of distributed ownership, exploring how retaining ownership, possession, and control of digital expression may enable alternative forms of creator compensation, exchange, and economic organization. Part~\ref{sec:conclusion} concludes.

This Article's central contribution is demonstrating that foundational copyright principles, applied within grassroots architectures, can support person-centered ownership rather than platform extraction. The argument does not depend on legal reform. Instead, it argues that existing copyright doctrine may produce different ownership outcomes when digital expression is created, possessed, and circulated within architectures that do not rely on third-party custody or contractual intermediation. 

If digital speech acts qualify for copyright protection, then copyright law may provide more than a mechanism for protecting software and platform business models; it may also furnish a legal foundation for alternative forms of digital organization grounded in distributed ownership, accountability, and democratic participation.
\section{Platform Governance Through Copyright and Contract}\label{sec:platform-regime}

\subsection{Historical Foundations}

The dominance of U.S.-based digital platforms over global Internet architecture is not happenstance, but the product of specific historical, regulatory, and legal conditions during the Internet's formative years. 

Under an economic paradigm termed "collective invention," the early development of the Internet took place almost entirely within the United States.\footnote{\label{fn:greenstein}Shane Greenstein, \textit{How the Internet Became Commercial: Innovation, Privatization, and the Birth of a New Network} (2015).} Essential protocols, including file transfer and TCP/IP, were developed in the early 1970s while the Internet (then "ARPANET") was still a research project owned by the U.S. Department of Defense.\footnote{Radu, \textit{supra} note~\ref{fn:radu}, at 44.} The principle of "end-to-end" design—placing computing power on users' devices rather than in network infrastructure—was articulated by Saltzer, Reed, and Clark in 1984\footnote{J.H. Saltzer, D.P. Reed \& D.D. Clark, \textit{End-to-End Arguments in System Design}, 2 ACM Transactions on Computer Systems 277 (1984).} and played an integral role during the Internet's stewardship by the U.S. National Science Foundation from 1985 to 1995.\footnote{Greenstein, \textit{supra} note~\ref{fn:greenstein}.} As this design principle guided the network's technical architecture, privatization—the transfer of publicly-owned Internet infrastructure to private actors who marketed and sold access to users—began in the late 1980s and accelerated through the early 1990s, transforming the nascent Internet from a U.S. national communication network to the dominant global network of data exchange it is today.\footnote{Julia Pohle \& Daniel Voelsen, \textit{Centrality and Power: The Struggle over the Techno-Political Configuration of the Internet and the Global Digital Order}, 14 Pol'y \& Internet 13 (2022).}

This privatization occurred within a distinctive regulatory environment. While the U.S. government had provided intensive funding and research support during the Internet's development as a public infrastructure, its approach shifted as privatization progressed. The government increasingly adopted permissive regulatory and policy approaches designed to promote the Internet's commercial proliferation, reducing direct state intervention and giving preferential treatment to market mechanisms.\footnote{Radu, \textit{supra} note~\ref{fn:radu}.} In 1988, the Federal Communications Commission created the category of "value-added services" for digital communications, leaving computer-mediated information virtually unregulated.\footnote{Amendment of Section 64.702 of the Commission's Rules and Regulations (Third Computer Inquiry), Report and Order, 104 F.C.C.2d 958 (1986), reconsidered, 2 F.C.C. Rcd. 3035 (1987), further reconsidered, 3 F.C.C. Rcd. 1135 (1988); see also Radu, \textit{supra} note~\ref{fn:radu}, at 45.} The Clinton Administration's 1996 "Framework for Global Electronic Commerce" cemented this approach, stressing self-regulation and private sector development of Internet operations to governments around the world.\footnote{William J. Clinton \& Albert Gore, Jr., A Framework for Global Electronic Commerce (1997), available at https://clintonwhitehouse4.archives.gov/WH/New/Commerce/}

Within this unique context, U.S. courts emerged at critical junctures in the Internet's development. While early technical standards and governance structures arose from quasi-private arrangements, the state's deliberate privatization and multi-stakeholder governance approach created an environment where private actors sought judicial mediation only on issues that could not be resolved privately. Federal courts became the forum for resolving disputes over domains, boundaries, rights, and obligations of private actors within Internet architecture—and, as Part II demonstrates, consistently resolved ambiguities in copyright law in ways that favored software developers over users.\footnote{See infra Part II.B–C.}

\subsection{Copyright: Code as Literary Works}

The legal foundation for platform copyright dominance begins with the Copyright Act of 1976 and its 1980 amendments addressing computer programs. Congress included computer programs within the definition of "literary work"—"works, other than audiovisual works, expressed in words, numbers, or other verbal or numerical symbols or indicia"—but only "to the extent that they incorporate authorship in the programmer's expression of original ideas as distinguished from the ideas themselves."\footnote{Lajos P. Pataki, Jr., \textit{Copyright Protection for Computer Programs Under the 1976 Copyright Act}, 52 Ind. L.J. 503, 504-05 (1977).} This qualified inclusion, however, left critical questions unresolved. Many aspects unique to computer code remained unaddressed: whether a program contained entirely within a computer's electronics was a "copy," whether temporary storage in a computer constituted making a "copy," and whether copyright law protects work that blends expression and utility.\footnote{\textit{Id.}}

To address these concerns, Congress created the National Commission on New Technological Uses of Copyrighted Works (CONTU) in 1974, which reviewed whether computer programs met copyrightability requirements and whether programs in machine-readable form were copyrightable material.\footnote{\label{fn:contu}National Commission on New Technological Uses of Copyrighted Works, \textit{Final Report of the National Commission on New Technological Uses of Copyrighted Works} 9 (1978).} CONTU's report, published in 1978 after the passage of the 1976 Copyright Act, led to recommendations that Congress adopted substantially in the 1980 amendments.\footnote{Pamela Samuelson, CONTU Revisited: The Case Against Copyright Protection for Computer Programs in Machine-Readable Form, 1984 Duke L.J. 663 (1984).} Critically, CONTU recognized it was impossible to draw a clear line between copyrightable "expression" and uncopyrightable "ideas" in computer programs—a distinction central to copyright law.\footnote{CONTU Final Report, \textit{supra} note~\ref{fn:contu}, at 22.} Drawing the line too far toward copyright protection would grant strong monopolies over specific applications; drawing it too far toward public domain would allow easy copying of programmers' work, discouraging creation.\footnote{\label{fn:menell}Peter S. Menell, \textit{An Analysis of the Scope of Copyright Protection for Application Programs}, 45 Stan. L. Rev. 1045 (1993).} Unable to resolve this tension, CONTU and Congress left this fundamental question to the courts.

The Third Circuit's decision in \textit{Apple Computer v. Franklin Computer} (1983) provided the answer: comprehensive copyright protection for all computer code, regardless of function, form, or medium. Franklin Computer copied portions of Apple's operating system software to create compatible computers, defending on grounds that purely functional code—especially operating systems in machine-readable object code stored in ROM—could not be copyrighted.\footnote{\textit{Apple Computer, Inc. v. Franklin Computer Corp.}, 545 F. Supp. 812 (E.D. Pa. 1982).} The Third Circuit rejected every aspect of this defense. Against the argument that object code (binary sequences of 0s and 1s unintelligible to humans) was uncopyrightable due to unintelligibility, the court held it was "the clear intention of Congress to protect all computer code."\footnote{\textit{Apple Computer, Inc. v. Franklin Computer Corp.}, 714 F.2d 1240, 1249 (3d Cir. 1983).} Against the argument that code was uncopyrightable because it had utilitarian purpose, the court held that Congress made "no distinction between application programs and operating programs."\footnote{\textit{Id.}} Against the argument that code stored in ROM was a mechanical device rather than a fixed expression, the court held that "fixation... is satisfied through the embodiment of the expression in the ROM devices."\footnote{\textit{Id.}}

The decision established comprehensive copyright protection for computer code and resolved critical questions Congress had left open. The Third Circuit held that all computer code receives copyright protection regardless of form—source code or object code—medium of storage—ROM chips, semiconductor devices, or magnetic disks—or function—application programs or operating systems.\footnote{\label{fn:nussbaum}Jeffrey L. Nussbaum, \textit{Apple Computer, Inc. v. Franklin Computer Corporation Puts the Byte Back into Copyright Protection for Computer Programs}, 14 Golden Gate U. L. Rev. 281, 281--82 (1984).} By rejecting Franklin's arguments that object code was too unintelligible for copyright, that code stored in ROM was a utilitarian object rather than fixed expression, and that operating systems were purely functional works outside copyright's scope, the court granted software manufacturers extensive exclusive rights over their entire code base.

Subsequent litigation refined the boundaries of this protection without challenging its core. Courts addressed whether reverse engineering for interoperability constituted fair use,\footnote{Sega Enters. Ltd. v. Accolade, Inc., 977 F.2d 1510 (9th Cir. 1992).} whether menu command hierarchies were copyrightable expression or uncopyrightable methods of operation,\footnote{\label{fn:lotus}
Lotus Dev. Corp. v. Borland Int'l, Inc., 49 F.3d 807 (1st Cir. 1995), aff'd by an equally divided court, 516 U.S. 233 (1996).} and where precisely to draw the idea/expression line for functional program elements. But the fundamental holding that all computer code qualifies for copyright protection regardless of form or function has remained settled law.

\subsection{The Copyright Paradox: Software Licensing}

Even before \textit{Apple v. Franklin} established comprehensive code protection, the software industry had intentionally developed a mechanism in the 1970s to circumvent copyright exemptions: the software license.\footnote{\label{fn:rustad}Michael L. Rustad, \textit{Software Licensing: Principles and Practical Strategies}, Suffolk U. L. Sch. Legal Studies Research Paper Series, Research Paper 14-5 (2014).} Concerned that competitors could purchase software copies and then lease them to consumers (who could duplicate them) or prepare derivative programs, software developers deliberately characterized transactions as licenses rather than sales.\footnote{\textit{Id.}} By making each license "personal and non-transferable," producers specifically targeted the first-sale doctrine—the copyright exemption allowing rightful owners of copyrighted material to sell, lend, or destroy the physical object without the copyright holder's consent.\footnote{\textit{Id.}}

This created the copyright paradox. Software licenses serve dual purposes: controlling the number of users, permitted locations, uses, and duration of software use,\footnote{Rustad, \textit{supra} note \ref{fn:rustad}.} while simultaneously transforming the enforcement mechanism for unauthorized use. By characterizing transactions as licenses rather than sales, any breach of license terms becomes a contract law violation rather than copyright infringement.\footnote{\label{fn:kim}Nancy S. Kim, \textit{The Software Licensing Dilemma}, 2008 BYU L. Rev. 1103 (2008).} Copyright law validates the license as an enforceable contract,\footnote{\label{fn:madison}Michael J. Madison, \textit{Reconstructing the Software License}, 35 Loy. U. Chi. L.J. 275 (2003).} but the license itself disenfranchises users from the protections copyright law provides—a structure that would later scale from individual software transactions to global platform governance.\footnote{\textit{Id.}}

The paradox was reinforced by \textit{MAI Systems v. Peak Computer} (1993), which established two critical precedents. First, the Ninth Circuit held that merely loading software into RAM—the most temporary form of computer storage—constitutes creating a "fixed" copy subject to copyright protection.\footnote{\textit{MAI Systems}, \textit{supra} note~\ref{fn:mai}} The court's reasoning was widely criticized by legal scholars as illogical: it acknowledged that RAM copies are "destroyed when a computer is turned off" yet held them sufficiently permanent to constitute copies under the Copyright Act.\footnote{Kathryn Levin, \textit{Intellectual Property Law - MAI v. Peak: Should Loading Operating System Software into RAM Constitute Copyright Infringement?}, 24 Golden Gate U. L. Rev. 649 (1994); Bridget J. Murphy, \textit{Loading Software into RAM Creates a Copy: MAI Systems Corp. v. Peak Computer Inc.}, 10 Santa Clara High Tech. L.J. 499 (1994).} Although Congress later amended Section 117 to permit temporary RAM copies for maintenance and repair purposes,\footnote{17 U.S.C. § 117(c) (1998).} the core holding that RAM loading constitutes fixation has remained. This extension of copyright protection granted software manufacturers "an almost patent-like monopoly over their computer code and software programs," extending exclusive rights at the expense of software users.\footnote{Theodore Arriola, \textit{Software Copyright Infringement Claims After MAI Systems v. Peak Computer}, 69 Wash. L. Rev. 405 (1994).}

Second, the court validated software licensing as a viable framework for software transactions. MAI needed only to prove copyright ownership over their code; the license agreement then controlled the relationship, characterizing software customers as "licensees" bound by all license terms, without the Section 117 exemptions granted to "owners."\footnote{Michael E. Johnson, \textit{The Uncertain Future of Computer Software Users' Rights in the Aftermath of MAI Systems}, 44 Duke L.J. 327 (1994).} This distinction—licensee versus owner—became central to platform power: users could be denied the rights that copyright law grants to owners simply through contractual characterization.

The final component of platform copyright dominance emerged with \textit{ProCD v. Zeidenberg} (1996), which validated "shrinkwrap" licenses—contracts enclosed within product packaging that become binding when users open the package. The Third Circuit had previously rejected such licenses in \textit{Step-Saver Data Systems v. Wyse Technology}, holding that terms presented after sale could not bind purchasers.\footnote{\textit{Step-Saver Data Sys., Inc. v. Wyse Tech.}, 939 F.2d 91 (3d Cir. 1991).} Matthew Zeidenberg purchased ProCD's SelectPhone software at retail, ignored the enclosed license restricting use to non-commercial purposes, and resold the database over the Internet.\footnote{\textit{ProCD}, supra note~\ref{fn:procd}} ProCD sued for breach of license terms.\footnote{\textit{Id.}} Zeidenberg defended on the ground that the license terms were enclosed inside the package and therefore could not bind him at the moment of purchase—the same reasoning the Third Circuit had accepted in \textit{Step-Saver}.

The Seventh Circuit, in an opinion by Judge Frank Easterbrook, held that shrinkwrap licenses are enforceable contracts under the Uniform Commercial Code, directly contradicting \textit{Step-Saver}.\footnote{\textit{Id.} at 1450.} Judge Easterbrook analogized shrinkwrap licenses to other standard-form contracts such as insurance policies and airline tickets, where "the exchange of money precedes the communication of detailed terms."\footnote{\textit{Id.} at 1449.} Merely providing the opportunity to read license terms was sufficient for enforceability: "Notice on the outside, terms on the inside, and a right to return the software for a refund if the terms are unacceptable... may be a means of doing business valuable to buyers and sellers alike."\footnote{\textit{Id.}} Dismissing concerns about unequal bargaining power, Judge Easterbrook wrote: "Competition among vendors, not judicial revision of a package's contents, is how consumers are protected in a market economy."\footnote{\textit{Id.} at 1451.} The court justified this conclusion on economic efficiency grounds: standard form contracts saved both consumers and producers time and money in negotiating individual contracts, and software consumers were "better off" under enforceable shrinkwrap licenses.\footnote{\label{fn:grusar}Brett L. Grusar, \textit{Contracting Beyond Copyright: ProCD, Inc. v. Zeidenberg}, 10 Harv. J.L. \& Tech. 353 (1997).}

\textit{ProCD} created a circuit split that opened the legal market for adhesion contract experimentation. With courts divided on enforceability, companies could draft Terms of Service and test their limits, knowing some jurisdictions would uphold them. Subsequent litigation did not overturn \textit{ProCD}'s core holding that online adhesion contracts can be enforceable. Instead, courts refined the boundaries of enforceability. The Second Circuit held in \textit{Specht v. Netscape} that browsewrap agreements fail when users lack sufficient notice of terms.\footnote{\textit{Specht v. Netscape Commc'ns Corp.}, 306 F.3d 17, 35 (2d Cir. 2002).} The District of Kansas held in \textit{Klocek v. Gateway} that terms inside a box are not part of the contract when the contract is formed at purchase.\footnote{\textit{Klocek v. Gateway, Inc.}, 104 F. Supp. 2d 1332, 1341 (D. Kan. 2000).} The Eastern District of Pennsylvania in \textit{Bragg v. Linden Research} found certain Terms of Service unconscionable.\footnote{\textit{Bragg v. Linden Research, Inc.}, 487 F. Supp. 2d 593, 607 (E.D. Pa. 2007).}

These decisions established limiting principles—conspicuous notice, meaningful opportunity to review, substantive unconscionability review—without rejecting adhesion contracts altogether. Platforms adapted by implementing mechanisms combining clickwrap (requiring explicit "I agree" clicks) and browsewrap (presenting terms via conspicuous hyperlinks), adding arbitration clauses to avoid judicial review of substantive terms, and structuring agreements to satisfy the notice requirements courts demanded.\footnote{See Nancy Kim, \textit{Wrap Contracts: Foundations and Ramifications} (2013).}

\subsection{Juridical Infrastructures}

\textit{Apple}, \textit{MAI}, and \textit{ProCD} did more than establish the specific doctrines for which they are commonly cited. Taken together, these decisions created a legal architecture that enabled new forms of governance through software and contracts. The cumulative effect is best understood as the construction of what this Article terms \emph{juridical infrastructures}.\footnote{\label{fn:juridical-infrastructures}
The concept of ``juridical infrastructures'' is developed more fully in James Golike,
\textit{The Copyright Paradox: How Three Court Cases Created Juridical Infrastructures
for Platform Governance} (on file with author).}

This Article conceptualizes juridical infrastructures as legal frameworks established through public legal institutions that function as underlying structures through which private actors exercise governance.
The concept combines insights from infrastructure studies and critical legal scholarship. Infrastructure scholars have demonstrated that infrastructures are not merely technical systems but socio-technical arrangements that enable and constrain particular forms of action while embedding relationships of power within seemingly neutral systems.\footnote{See Susan Leigh Star, \textit{The Ethnography of Infrastructure}, 43 Am. Behav. Scientist 377 (1999); Jean-Christophe Plantin et al., \textit{Infrastructure Studies Meet Platform Studies in the Age of Google and Facebook}, 20 New Media \& Soc'y 293 (2018).} Critical legal scholarship similarly observes that law does not merely regulate social relations but constitutes the conditions under which particular social, economic, and political arrangements become possible.\footnote{See Julie E. Cohen, \textit{Between Truth and Power: The Legal Constructions of Informational Capitalism} (2019); Alan Hunt, \textit{The Concept of Law and the Everyday World}, in \textsc{Reading Dworkin Critically} 35 (1992).}

Juridical infrastructures combine these insights. Just as technical infrastructures shape what actions can occur within a system, legal arrangements allocate rights, obligations, and authority, determining who may participate, on what terms, and subject to whose control. Copyright doctrines, licensing arrangements, contractual forms, and judicial precedents are therefore not merely rules governing platforms. They are part of the operative infrastructure through which platform governance itself becomes possible.

The interaction of \textit{Apple}, \textit{MAI}, and \textit{ProCD} illustrates this process. \textit{Apple} established broad copyright protection for software. \textit{MAI} transformed software use into licensed access rather than ownership. \textit{ProCD} validated contractual restrictions imposed through mass-market licenses. None of these decisions created platform governance independently. Together, however, they generated the legal conditions that allowed platform firms to govern users through proprietary software, licensing arrangements, and contractual terms. Even as courts refined the boundaries and limitations of these doctrines through later litigation, what emerged was not merely a collection of decisions but a durable juridical infrastructure capable of supporting platform governance at global scale.

The significance of juridical infrastructures is not merely explanatory. Once established, they become operationalized through concrete institutional arrangements. In the platform economy, the principal mechanism through which these infrastructures are deployed is the Terms of Service agreement.

\subsection{From Software Licensing to Platform Terms of Service}

Software licensing transformed the role of contract within digital environments. As software licences became the primary mechanism through which users accessed copyrighted works, they ceased to function merely as bilateral agreements between software developers and individual consumers. Instead, they increasingly operated as systems of private ordering that governed all users of a particular software environment, establishing uniform rules for participation applicable to every participant.\footnote{See Madison, \textit{supra} note~\ref{fn:madison}, at 288--94 (describing software licensing as a form of private lawmaking).}

The transition from individual software licences to platform Terms of Service extended this logic to online platforms. Terms of Service no longer governed only individual access to software; they governed participation in entire communicative and economic environments, allocating rights, obligations, and authority among billions of users through privately drafted contractual terms. The legal framework established by \textit{Apple v. Franklin}, \textit{MAI v. Peak}, and \textit{ProCD} therefore did not remain confined to software transactions. As software migrated from physical media to internet-based services, the licensing model evolved into a governance framework capable of organizing participation at platform scale.

Every major platform operationalizes this approach through its Terms of Service. Facebook's Terms of Service state: "If you do not agree to these Terms, then do not access or use Facebook or the other products and services covered by these Terms."\footnote{Meta, Terms of Service, \textit{supra} note 1} YouTube provides: "If you do not understand the Agreement, or do not accept any part of it, then you may not use the Service."\footnote{YouTube, Terms of Service, \textit{supra} note 1.} X declares: "By using the Services you agree to be bound by these Terms."\footnote{X, Terms of Service, \textit{supra} note 1.} These agreements are offered in pure "take-it-or-leave-it" structure: users agree by entering the platform, regardless of whether they have read or understood any terms.\footnote{Grusar, \textit{supra} note~\ref{fn:grusar}.}

While these provisions govern who may participate in platform environments, the substantive allocation of rights and authority occurs through the licensing arrangements embedded within platform Terms of Service. 

At the core of every platform Terms of Service is a licensing exchange that structures the legal relationship between platform and user. Platforms grant users a "personal, worldwide, royalty-free, non-assignable and non-exclusive license to use the software" provided by the platform.\footnote{YouTube, Terms of Service, \textit{supra} note 1} This license is revocable—platforms can terminate user accounts at any time, for any reason, often without explanation. In exchange, users grant platforms comprehensive licenses to user-generated content. Facebook requires users to grant a "non-exclusive, transferable, sub-licensable, royalty-free and worldwide licence to host, use, distribute, modify, run, copy, publicly perform or display, translate and create derivative works of" all user content.\footnote{Meta, Terms of Service, \textit{supra} note 1} YouTube requires a "worldwide, non-exclusive, royalty-free, transferable, sublicensable licence to use that Content (including to reproduce, distribute, modify, display and perform it)" for operating, promoting, and improving the service, plus the "right to monetize" user content.\footnote{YouTube, Terms of Service, \textit{supra} note 1} X requires a "worldwide, non-exclusive, royalty-free license (with the right to sublicense) to use, copy, reproduce, process, adapt, modify, publish, transmit, display, upload, download, and distribute" user content "in any and all media or distribution methods now known or later developed, for any purpose," including to "analyze text and other information" and let others do the same, "with no compensation paid."\footnote{X, Terms of Service, \textit{supra} note 1}

\subsection{The Economics of the Platform License Exchange}

Platforms assert comprehensive copyright protection over every line of code in their ecosystems—from operating systems to APIs—through the precedents of \textit{Apple v. Franklin}. Users are characterized as "licensees" rather than "owners" through the precedent of \textit{MAI v. Peak}, denying them the rights copyright law grants to owners. The license exchange is validated through the precedent of \textit{ProCD}, despite the complete absence of competitive markets that supposedly justify adhesion contracts. Users nominally retain copyright in content they create, but platforms obtain such broad licenses—worldwide, sublicensable, royalty-free, for any purpose including commercial use and AI training—that nominal ownership becomes practically meaningless. X makes the extraction explicit: "the use of the Services by you is hereby agreed as being sufficient compensation for the Content and grant of rights herein."\footnote{X, Terms of Service, \textit{supra} note 1} Platform access itself becomes payment, transforming user expressions into platform capital without direct monetary exchange. YouTube's Terms state bluntly: "This Agreement does not entitle you to any payments" from monetization of user content.\footnote{YouTube, Terms of Service, \textit{supra} note 1}

This contractual structure enables platforms to accumulate vast wealth extracted from user-generated value. Meta reported revenues of \$200.97 billion in 2025, generating ``substantially all'' of its revenue from advertising based on user data and content.\footnote{\label{fn:meta2025}
Meta Platforms, Inc., Annual Report (Form 10-K) for fiscal year ended Dec. 31, 2025, at 15, 60 (filed Jan. 30, 2026), https://www.sec.gov/ix?doc=/Archives/edgar/data/0001326801/000162828026003942/meta-20251231.htm [hereinafter Meta 2025 Form 10-K].}Alphabet (Google/YouTube) generated \$402.8 billion the same year, with ``more than 70\%'' from online advertising systems that similarly monetize user-generated content, searches, and engagement data without compensation to users.\footnote{Alphabet Inc., Annual Report (Form 10-K) for fiscal year ended Dec. 31, 2025, at 9, 32 (filed Jan. 31, 2026), https://www.sec.gov/ix?doc=/Archives/edgar/data/0001652044/000165204426000018/goog-20251231.htm.} This extraction occurs through Terms of Service agreements that grant platforms comprehensive control over access, use, and interaction within digital spaces while compelling users to surrender licensing rights to the data and content that generates this wealth.

The economic implications of the platform license exchange extend beyond the licensing provisions themselves. The broader contractual structure governing platform participation preserves extensive platform discretion while limiting users' ability to influence the terms under which value is created, governed, and captured. Facebook's Terms significantly omit any explicit software license grant to users, suggesting a shift from licensing contracts toward pure property-based governance where platform access becomes a revocable privilege rather than a contractual right.\footnote{Meta, Terms of Service, \textit{supra} note 1.} The variation in notification periods for Terms of Service changes reveals a similar asymmetry. Facebook commits to "at least 30 days" notice before changes; YouTube offers only "reasonable advance notice"; X merely states "We will try to notify you of material revisions."\footnote{Meta, Terms of Service \S~4.1, \textit{supra} note 1; YouTube, Terms of Service, \textit{supra} note 1; X, Terms of Service, \textit{supra} note 1.} Users have no ability to negotiate, debate, or refuse such changes—continued use constitutes acceptance.

\subsection{Ownership and Possession Separated}

Beyond the asymmetries embedded in the platform license exchange, platform architecture creates a second asymmetry: the separation of ownership and possession. Users in principle own copyright in the content they create, but the centralized server-based infrastructure ensures platforms physically retain the copyrighted expression. The person holds legal rights under 17 U.S.C. § 106; the platform holds the data and controls access through both Terms of Service and platform technical infrastructure. 

The platform exploits this separation to exercise what legal scholars have characterized as infrastructural power over users and their data.\footnote{See \label{fn:cohen-networked}Julie E. Cohen, \textsc{Configuring the Networked Self: Law, Code, and the Play of Everyday Practice} 122--52 (2012) (analyzing how platform architecture separates user ownership from platform control); Balkin, \textit{supra} note~\ref{fn:balkin-fiduciaries}, at 1186--89 (platforms as fiduciaries controlling user information); Pasquale, \textit{supra} note~\ref{fn:pasquale-blackbox}, at 19--58 (opacity of platform control over user data).} Platforms obtain personal information, social graph data, and communication content; commercially exploit them through what Zuboff terms "surveillance capitalism";\footnote{Shoshana Zuboff, \textit{The Age of Surveillance Capitalism: The Fight for a Human Future at the New Frontier of Power} (2019).} and render user copyright ownership effectively meaningless because access to one's own data depends on platform discretion. This separation of ownership from possession—users retain legal rights while platforms exercise physical and technical control—enables the extraction that Terms of Service agreements legitimate.

\subsection{The Limitations of Regulatory Responses}

Regulatory responses to platform dominance have addressed data protection and content governance without challenging the underlying contractual and technical framework that enables extraction. The European Union's GDPR imposes consent requirements, data portability rights, and privacy protections.\footnote{Regulation (EU) 2016/679 of the European Parliament and of the Council of 27 April 2016 on the protection of natural persons with regard to the processing of personal data (General Data Protection Regulation).} The Digital Services Act requires content moderation transparency and appeals mechanisms.\footnote{\label{fn:dsa}Regulation (EU) 2022/2065 of the European Parliament and of the Council of 19 October 2022 on a Single Market For Digital Services (Digital Services Act).} These regulatory interventions have achieved meaningful protections: GDPR's consent requirements and right to erasure provide users some control over personal data processing,\footnote{See Daniel J. Solove \& Paul M. Schwartz, \textit{Privacy Law Fundamentals} 115--47 (2021) (analyzing GDPR's effectiveness in constraining data processing).} and the Digital Services Act's transparency obligations improve accountability for content moderation decisions.\footnote{See Daphne Keller, \textit{Empirical Evidence of Over-Removal by Internet Companies Under Intermediary Liability Laws}, Stanford Ctr. for Internet \& Soc'y (2015) (analyzing platform moderation under various regulatory regimes).}

Yet these protections leave the fundamental copyright appropriation structure intact. Platforms continue to require users to grant comprehensive content licenses as a condition of access. Users continue to surrender possession of their data and content to platform servers. Platforms continue to extract commercial value from user-generated content without compensation—GDPR does not prevent Meta from requiring users to grant "worldwide, non-exclusive, transferable, sub-licensable, royalty-free" licenses to everything they create, nor does the Digital Services Act prevent platforms from monetizing user content through advertising. Privacy protections and content moderation rules address governance of data already surrendered, not the initial transfer of possession and licensing of rights.

The distinction is therefore not between effective and ineffective regulation, but between regulatory and architectural responses to platform power. GDPR and the Digital Services Act address how platforms process, manage, and moderate data and content after those resources have entered intermediary-controlled systems. The power asymmetry identified in this Article arises earlier: from the separation of ownership and possession enabled by contract and copyright law and operationalized through platform architecture. Regulatory protections operate downstream from that separation, constraining certain platform practices without altering the underlying conditions through which ownership and possession become divided. The architectural intervention examined here operates upstream, preventing the separation of ownership and possession by design and thereby reducing dependence on regulatory protections for data already placed under intermediary control.

\subsection{Alternative Architectural Arrangements}

The preceding analysis demonstrates that contemporary platform dominance is not simply a consequence of technological innovation. It emerged through the interaction of copyright doctrine, software licensing, contractual ordering, and architectural design. The legal categories developed during the software era—including ownership, licensing, and user status—were subsequently incorporated into platform architectures and applied at global scale.

Recognizing the historical contingency of these arrangements, however, reveals that alternative configurations are possible. If ownership, possession, and authority are distributed through particular combinations of law and architecture, different architectural arrangements may produce different distributions of control and participation.

This Article examines one such arrangement. Rather than relying upon centralized platforms that aggregate possession of data and govern participation through intermediary-controlled infrastructure, grassroots architectures distribute possession among participants themselves. Individuals create, store, and exchange digitally authenticated expressions through devices they control. As a result, they present a different relationship between authors,
intermediaries, and digital participation than that found in contemporary platform environments.
\section{Grassroots Platforms: Architectural Prevention of Copyright Extraction}\label{sec:grassroots}

Grassroots architectures are serverless, permissionless platforms that operate solely on the networked smartphones of their cryptographically identified participants, enabling individuals to retain possession and control over their personal information.\footnote{Shapiro, \textit{supra} note~\ref{fn:shapiro-grassroots-systems}.}

\subsection{Volitional and Machine Transactions}

The formal framework for grassroots platforms uses \textit{volitional multiagent atomic transactions}.\footnote{Lewis-Pye \& Shapiro, \textit{supra} note~\ref{note:vmat}.}  Each agent's state has two components: \textit{volitions}---what the person wants---and \textit{machine state}---the state of their device.  Transactions are of two kinds: \textit{volition transactions}, in which a single person freely changes their volitions, and \textit{machine transactions}, which change only machine state and are guarded by conditions on the volitions of the participating agents.

Creating a digital speech act---deliberately signing content---is a volition transaction: a free, unary act by a single person, with no precondition.  The person decides to sign specific content; this decision is theirs alone.

The digital social contract then specifies machine transactions that are guarded by the presence of a digital speech act.  Two such rules are central to copyright:

\textbf{Non-unbundling.}  When a digital speech act is forwarded, the signed content cannot be separated from the signature.  The digital social contract enforces this as a machine transaction: forwarding copyrighted content requires embedding the original digital speech act---signature and all---within the forwarder's new digital speech act.  A recipient who receives copyrighted content receives the creator's signature with it, and cannot strip it.

\textbf{Provenance preservation.}  When a digital speech act is forwarded, the full provenance chain---the sequence of forwarders---must be included.  Each forwarder's signature is added to the chain, ensuring that every person who forwarded the content is identified, establishing accountability at each step.

Together, non-unbundling and provenance preservation are the two rules by which the digital social contract enforces copyright: the creator's claim cannot be stripped, and the full history of who forwarded the content is maintained.

\subsection{Enforcement via Mutual Attestation}

A digital social contract must be enforced: participants must not be able to deviate from the contract as programmed.  For grassroots platforms, enforcement is achieved through \textit{mutual attestation}: when two agents communicate, each proves to the other that the message was produced by a correct execution of the agreed-upon contract code.  An attested agent either executes correctly or crashes---it cannot behave arbitrarily.\footnote{\label{note:cssn}Secure GLP (Ehud Shapiro, personal communication, 2026).}

For digital social contracts that require no consensus---including social graphs and social networking---mutual attestation among the participants suffices for enforcement.  The contract runs on the participants' own devices, and compliance is verified bilaterally.  No third-party validators, no global consensus protocol, and no blockchain are required.\footnote{Shapiro, \textit{supra} note~\ref{note:cssn}.}

\subsection{Multiple Independent Instances}

A grassroots platform can have multiple instances that emerge and operate independently of each other and of any global resource except the network, and can interoperate and coalesce once interconnected, potentially forming a single global instance.\footnote{\label{note:icdcn}Ehud Shapiro, \textit{Grassroots Platforms with Atomic Transactions: Social Networks, Cryptocurrencies, and Democratic Federations}, in Proceedings of the 27th International Conference on Distributed Computing and Networking (ICDCN 2026) 71--81 (2026) (also at arXiv:2502.11299).}  This is in contrast to global platforms, which can only have a single instance---one Facebook, one Bitcoin---and to federated systems such as Mastodon, where each instance depends on its operator's server.\footnote{Aravindh Raman et al., \textit{Challenges in the Decentralised Web: The Mastodon Case}, in Proceedings of the Internet Measurement Conference 217--222 (2019).}

The grassroots property ensures that any group of people with smartphones can form their own platform instance, independently of any existing instance or global resource.  Independently-formed instances may interconnect---for example, when a person in one instance befriends a person in another---and may eventually coalesce into a larger instance.  At no point does any third party obtain control over the platform or the digital speech acts of its participants.

\subsection{No External Legal Entity}

In a grassroots platform, there are no servers, no platform operators, no data collection, no algorithmic content selection, and no advertising.\footnote{Shapiro, \textit{supra} note~\ref{note:gsn}.}  Each person stores only the local neighbourhood pertaining to them on their own smartphone, and no third party has access unless explicitly granted.

The absence of an operating entity has a direct legal consequence: there is no legal person that can draft Terms of Service, impose license agreements, or claim copyright ownership over the expressions of the platform's participants.  On existing platforms, the entity that operates the platform is also the entity that imposes Terms of Service requiring people to grant broad copyright licenses as a condition of participation.  In a grassroots platform, no such entity exists.  An app store that distributes the software imposes terms on the software, not on the content a person creates.  The software can also be distributed independently of any such store.

\section{Digital Speech Acts: Technical Definition and Legal Foundation}\label{sec:dsa}

Cardelli et al.\ introduced the term ``digital speech act'' to describe a cryptographically-signed, indexed transaction (an ``utterance'') within a digital social contract---a voluntary agreement among a set of people, specified, fulfilled, and enforced digitally.\footnote{Cardelli et al., \textit{supra} note~\ref{note:cardelli}.} In its original formulation, a digital speech act by an agent consists of signing an utterance with the agent's private key and broadcasting it to the other parties of the contract. Cardelli et al.\ focus on digital speech acts as primitives for constructing egalitarian social contracts through computational mechanisms.

This Article extends the concept in three directions relevant to digital self-governance. First, we ground digital speech acts in both speech act theory (Searle and Austin) and volitional multiagent atomic transactions (VMAT),\footnote{See infra Part IV.D.} establishing that digital speech acts are deliberate performative utterances, not merely cryptographic operations. Second, we demonstrate that digital speech acts should qualify for copyright protection under existing U.S. precedent,\footnote{See infra Part V.} providing legal enforceability beyond the computational guarantees Cardelli et al.\ analyze. Third, we connect the legal status of digital speech acts, including the copyright interests they may generate, to political theory on property and democratic sovereignty,\footnote{See infra Parts~\ref{sec:ownership-sovereignty} and~\ref{sec:democratic-foundations}.} establishing that recognition of such rights is an important prerequisite for digital self-governance. These extensions transform digital speech acts from computational primitives into legal objects with both technical and normative force.

Digital speech acts arise in many contexts: transactions in grassroots currencies and bonds\footnote{Ehud Shapiro, \textit{Grassroots Currencies: Foundations for Grassroots Digital Economies} (2024) (arXiv preprint arXiv:2202.05619).} are digital speech acts, as is voting in democratic digital communities.\footnote{Ehud Shapiro \& Nimrod Talmon, \textit{Grassroots Federation: Fair Governance of Large-Scale, Decentralized, Sovereign Digital Communities} (2025) (arXiv preprint arXiv:2505.02208).} This Article focuses on expressive digital speech acts within grassroots social networks, which provide the principal context for the copyright analysis developed below.

\subsection{Technical Definition}

A \textit{digital speech act}, as developed here, is a deliberate volitional act by a person, performative in the sense of Searle and Austin, realized by cryptographically signing personal content---text, audio, photo, video, or structured data---with the person's private key. By performing this act, the person simultaneously establishes (\ia) attribution---the content is verifiably theirs; (\ib) accountability---they stand behind it and can be held responsible for it; and (\ic) authorship---they are the creator of the resulting expression. A digital speech act exists independently of any particular social contract or platform; it is an act of a person, carried out on the person's own device.

\textbf{Digital speech acts} are created through the combination of three elements: a self-chosen cryptographic key pair that uniquely identifies the person, content created or selected by the person (text, audio, photo, video, or structured data), and the cryptographic signature produced by signing that content with the person's private key. The key pair is generated by the person on their own device. Content may include previously received digital speech acts being forwarded, in which case the forwarded speech act is embedded within the new one. The signature produces a unique expression verifiable by any third party using the person's public key. The signed data includes a timestamp recording when the act was performed.

These three elements combine to produce a computationally unique cryptographic string. This string is a fixed expression that cannot be produced by anyone lacking the private key and can be verified by anyone with access to the public key and signed content. A digital speech act exists independently of any particular social contract or platform; it is created by a person on their own device and can be stored, transmitted, or published through any medium while retaining its cryptographic properties.

The distinction between digital speech acts and superficially similar cryptographic operations becomes crucial for copyright analysis. Consider the difference between a digital speech act and a platform authentication token. A person creating a digital speech act deliberately chooses content---``I certify that I will deliver quality work by March 15, 2026''---and signs it, combining authentication function with expressive content. A platform authentication token, by contrast, is generated automatically by platform systems without user content selection: ``User ID 12345 authenticated at 14:32:07 GMT.'' The authentication token serves pure function without expression; the digital speech act combines function (authentication) with deliberate expression (chosen content). This distinction matters because copyright protects expression, not mere function.

\subsection{Properties}\label{subsec:dsa-properties}

The technical architecture of digital speech acts produces five properties relevant to their function and legal status within grassroots platforms. 

First, digital speech acts are \textit{independently created}: each person generates their own content and data, along with a key pair on their own device, without reliance on central authority or trusted third party, maintaining autonomous control over key generation. Second, they are \textit{computationally unique}: cryptographic hash functions ensure that different persons produce distinct signatures even for identical content, and the same person produces different signatures for different content, making each signature unique with overwhelming probability to the combination of signer and content.

Third, digital speech acts are \textit{permanently fixed}: once created and signed, the cryptographic signature binds content to identity immutably---any content alteration invalidates the signature and is immediately detectable. Fourth, they are \textit{verifiably attributable}: attribution is cryptographically intrinsic, allowing anyone with the public key to verify the signature was produced by the holder of the corresponding private key, without dependence on platform operators, centralized databases, or intermediary trust. Fifth, they enable \textit{provenance preservation}: when a digital speech act is forwarded, the original is embedded as content within the forwarder's new digital speech act, creating cryptographic provenance chains that preserve attribution at each step and establish accountability for both original creators and subsequent forwarders.

\subsection{Two Parallel Foundations}

A digital speech act has two complementary characterizations: as a \textit{performative utterance} in the sense of Searle and Austin, and as a \textit{volitional transaction} in the sense of the volitional multiagent atomic transactions framework.\footnote{Lewis-Pye \& Shapiro, \textit{supra} note~\ref{note:vmat}.}  These are not competing accounts but parallel foundations: speech act theory provides the linguistic-philosophical grounding establishing that the act is expressive; the volitional-transactions framework provides the formal grounding establishing that the act is a deliberate volition by a person, distinct from machine state.

\subsubsection{Speech Act Theory: Performative Utterances (Searle and Austin)}

Searle's systematic taxonomy classifies illocutionary acts---utterances characterized by their illocutionary point (what the speaker intends to accomplish)---into five categories.\footnote{\label{note:searle}John R.\ Searle, \textsc{Expression and Meaning: Studies in the Theory of Speech Acts} (1979).} Declarations bring about change in reality through the utterance itself (``I hereby transfer three blue coins to Sue''; ``I vote YES on Proposal \#7''). Commissives commit the speaker to future action (``I promise to deliver by March 15''; ``I pledge to uphold these standards''). Assertives commit the speaker to the truth of a proposition (``I certify this document is authentic''; ``I confirm receipt of payment''). Directives attempt to get the addressee to do something (``I request membership in this community''; ``I invite Alice to join this group''). Expressives express the speaker's psychological state (``I approve this proposal''; ``I thank Bob for his contribution'').\footnote{For comprehensive treatment of speech act categories, see John R. Searle, \textit{A Taxonomy of Illocutionary Acts}, in \textsc{Expression and Meaning} 1--29 (1979).}

Digital speech acts can instantiate any of these categories, and creating one requires substantive creative choices. The person selects the propositional content (what to say: "I vote YES on Proposal \#7" versus "I vote NO"), the timing (when to perform the act), the identity (with which key pair to sign, when multiple are available), and the structural arrangement (how to format and structure the signed data). These layered dimensions of choice establish that creating a digital speech act entails expressive decisions rather than merely executing a functional operation. The person's choice of what to say determines the illocutionary force that results---voting, promising, certifying---but the force is a consequence of the content chosen, not a separate selection.

This taxonomy builds on Austin's foundational distinction between \textit{constative} utterances, which describe states of affairs and can be evaluated as true or false, and \textit{performative} utterances, which \textit{do things} through being uttered rather than merely describing.\footnote{J.L.\ Austin, \textsc{How to Do Things with Words} (1962).} Digital speech acts are performative: cryptographically signing ``I hereby transfer three blue coins to Sue'' does not describe a transfer, but \textit{performs} it.\footnote{Cardelli et al., \textit{supra} note~\ref{note:cardelli}.} 

The application of speech act theory to cryptographic operations is contested---some scholars question whether computational processes qualify as performative utterances when execution depends on algorithms rather than social convention.\footnote{See, e.g., Mireille Hildebrandt, \textit{Smart Technologies and the End(s) of Law: Novel Entanglements of Law and Technology} 117--45 (2015) (questioning whether smart contracts constitute performative speech); Tiziana Terranova, \textit{Red Stack Attack! Algorithms, Capital, and the Automation of the Common}, in \textsc{Accelerate: The Accelerationist Reader} 379--99 (Robin Mackay \& Armen Avanessian eds., 2014) (analyzing performativity in computational systems).} Digital speech acts differ: the person makes substantive choices about content, timing, and structure, and the cryptographic signature authenticates human agency. The performative force derives from deliberate human choice, not autonomous code execution.

\subsubsection{Volitional Multi-Agent Atomic Transactions (VMAT)}

In the volitional multiagent atomic transactions framework, each agent's state has two components: \textit{volitions}---what the person wants---and \textit{machine state}---the state of the device.\footnote{Lewis-Pye \& Shapiro, \textit{supra} note~\ref{note:vmat}.}  A \textit{volition transaction} is a free, unary act in which a single person changes their own volitions; it has no precondition.

Creating a digital speech act is a volition transaction: the person decides to sign specific content, and this decision is theirs alone. The cryptographic signing is the mechanism by which the volition becomes verifiable, fixed, and attributable, but the decision precedes and grounds the mechanism.

\subsection{Digital Speech Acts as Legal Objects}

Digital speech acts possess the characteristics copyright doctrine has historically associated with protectable expression: human authorship, original expression, fixation, and attribution. The question is therefore not whether copyright law has a framework capable of evaluating such expressions, but whether existing doctrine properly applied encompasses them.
\section{Legal Framework: Digital Speech Acts as Copyrightable Expression}\label{sec:core}

Copyright protection requires that works be original works of authorship fixed in a tangible medium of expression.\footnote{17 U.S.C. § 102(a).} Digital speech acts, as cryptographically signed expressions combining content, identity, and timestamp, created through deliberate personal choices and stored on personal devices, satisfy each element. \textit{Burrow-Giles v. Sarony} provides the primary doctrine for authorship: personal creative choices constitute authorship despite mechanical or algorithmic processes. \textit{Feist v. Rural} establishes the minimal creativity threshold for originality that digital speech acts easily satisfy. The Copyright Act's fixation requirement is clearly met by persistent storage on personal devices.

\subsection{What Copyright Protects in Digital Speech Acts}

Copyright protects expression, not ideas. 17 U.S.C. §102(b) excludes "any idea, procedure, process, system, method of operation" from protection. This  distinction determines what aspects of digital speech acts qualify for copyright: the \textbf{idea}—authentication, attribution, cryptographic identity verification—is unprotectable. These are methods of operation that §102(b) deliberately excludes. The \textbf{expression}—the digital speech act as a fixed, attributable work embodying specific creative choices—is protectable. The copyrightable work is the expressive content as embodied in the digital speech act—the signed, fixed, and attributable expression resulting from a person's deliberate creative choices.

Copyright does \textbf{not} protect the bare cryptographic signature string considered in isolation. For given inputs (content, private key, algorithm, timestamp), the signature output is computationally determined—there is only one signature that can result from those specific inputs. Claiming copyright in this deterministic output would raise merger concerns: if the idea is "authenticate this specific content with this specific key" and the expression is "this unique signature string," idea and expression may merge.

Copyright \textbf{does} protect the digital speech act—the deliberate act of communicating content through a cryptographically signed expression that becomes fixed and attributable. The underlying content may already qualify for copyright protection independently. The act of signing does not create copyright where none exists, nor does the inclusion of timestamps, signatures, or other provenance information defeat protection merely because those elements may be functional or factual. Instead, the digital speech act preserves and authenticates the author's expressive choices—what was expressed, by whom, and when—within a verifiable record. Copyright therefore protects the resulting expression as fixed, attributable, and embodied within a verifiable communicative act, while cryptographic mechanisms serve to preserve authorship and provenance rather than replace the underlying expressive content.

Not all digital speech acts qualify for, require, or benefit from copyright protection. Votes, token transfers, bare records, and similar functional grassroots transactions derive their legal and operational force from other sources. Digital signatures authenticate them, contract law enforces them, and the digital social contract governs their treatment within grassroots networks. Copyright is not the appropriate legal instrument for those objects, and the argument advanced here does not depend upon treating them as copyrightable works.

The copyright claim developed here is therefore limited to expressive digital speech acts whose content embodies the creative choices that copyright doctrine has traditionally protected. Text, audio, photographs, video, and similar forms of expression remain eligible for protection even when fixed, authenticated, and attributed through cryptographic means.

This conclusion is also consistent with longstanding copyright practice on centralized platforms. Expressive user-generated content posted to centralized platforms routinely receives copyright protection notwithstanding that it exists within platform-defined authentication systems, metadata structures, timestamps, and account architectures over which users exercise no creative control. No court has suggested that these technical containers defeat copyright in the underlying work. Grassroots architectures present the same analytical structure. Cryptographic signatures, provenance information, and protocol-defined metadata function as the technical infrastructure through which expression is authenticated and attributed, not as substitutes for the human creative choices embodied in the resulting work.

\subsection{Authorship under Burrow-Giles v.\ Sarony (1884)}

\textit{Burrow-Giles Lithographic Co. v. Sarony}, 111 U.S. 53 (1884), was the first Supreme Court case addressing whether a new technology could produce copyrightable works of authorship. Photography had existed for decades, but its mechanical nature raised fundamental questions: could photographs embody human creativity, or were they merely automatic reproductions produced by cameras rather than human authors?

The Court’s resolution established an analytical framework for evaluating whether technologies combining mechanical processes with human input produce copyrightable works—a framework courts have repeatedly adapted to new technologies ever since.\footnote{See, e.g., \textit{Goldstein v. California}, 412 U.S. 546 (1973) (sound recordings); \textit{Sony Corp. of Am. v. Universal City Studios, Inc.}, 464 U.S. 417 (1984) (home video recording technology); \label{fn:cablevision}\textit{Cartoon Network LP, LLLP v. CSC Holdings, Inc.}, 536 F.3d 121 (2d Cir. 2008) (remote digital video recorders); \textit{Authors Guild, Inc. v. Google, Inc.}, 804 F.3d 202 (2d Cir. 2015) (mass digitization and search technologies).} Under the \textit{Burrow-Giles} framework, the determinative inquiry is not whether mechanical processes are present—obviously they are in photography—but whether the work embodies "originality, of intellectual production, of thought, and conception on the part of the author."\footnote{\textit{Id.} at 60.} The Court defined "author" constitutionally as "he to whom anything owes its origin; originator; maker."\footnote{\textit{Id.} at 58.} Applying this framework, the Court held that photographs can be copyrighted when they embody the photographer's "original intellectual conceptions" through creative choices.

Napoleon Sarony's photograph of Oscar Wilde was copyrightable because Sarony "made the same... entirely from his own original mental conception, to which he gave visible form by posing the said Oscar Wilde in front of the camera, selecting and arranging the costume, draperies, and other various accessories... arranging the subject so as to present graceful outlines, arranging and disposing the light and shade, suggesting and evoking the desired expression."\footnote{\textit{Id.} at 60.} The mechanical-chemical processes of the camera preserved Sarony's creative choices in fixed form. The presence of mechanical execution did not negate authorship; what mattered was that the resulting work embodied deliberate human creative choices.

Digital speech acts present the same question \textit{Burrow-Giles} answered for photography: can a new technology combining algorithmic processes with deliberate human choices produce copyrightable works? We argue the answer follows this framework.

Applying \textit{Burrow-Giles} to digital speech acts, however, requires careful articulation because of a critical difference in how creative choices are embodied. Sarony's creative choices were perceptibly embodied in the photograph itself---a viewer could see the pose, lighting, and costume that Sarony selected. The creative choices were visible in the work. In contrast, a bare cryptographic signature string does not perceptibly embody timing choices or illocutionary force---the string does not "look like" a choice to sign now rather than later.

This disanalogy, however, does not defeat copyright in digital speech acts because the copyrightable work is not the bare signature string considered in isolation, but the digital speech act as a whole: the combination of content, cryptographic signature, and associated metadata (timestamp, public key identifier, signed data structure). This complete expression perceptibly embodies the creator's choices.

Content embodies expressive choices. The text, image, audio, or video that the person chooses to sign contains and displays their expressive choices: what they chose to say, how they chose to say it, the illocutionary force they chose to convey. These choices are perceptible in the content itself, just as Sarony's choices were perceptible in the photograph.

The signature preserves attribution and timing. The cryptographic signature does not exist to embody choices perceptibly to human viewers---it exists to make attribution and timing cryptographically verifiable. When a digital speech act is displayed or transmitted, the viewer sees: (1) the content (embodying what was said), (2) the signer's identity (embodying who said it), (3) the timestamp (embodying when it was said), and (4) the cryptographic signature that makes (2) and (3) verifiable rather than merely claimed. The signature's function is preservation and verification of attribution, not perceptible embodiment of aesthetic choices.

Together, content plus signature plus metadata form a complete work that perceptibly embodies the creator's choices about what to say, when to say it, and with which identity to claim it. The signature component serves the same function in digital expression that Sarony's signature on the photograph served: it establishes authorship and makes the attribution verifiable. The difference is that cryptographic signatures provide mathematical rather than visual verification. Yet \textit{Burrow-Giles} grounded authorship not in any particular form of attribution, but in whether the work embodied the author's original intellectual conceptions.

The parallel to digital speech acts is direct: just as Sarony's creative choices constituted authorship despite the camera's mechanical operation, a person's volitional and expressive choices constitute authorship despite cryptography's algorithmic operation. The constitutional definition of ``author'' therefore applies naturally to digital speech acts. The person who creates the content, chooses to sign it, and binds it to a cryptographic identity originates the work; the cryptographic system merely executes and records those choices.

The objection that cryptographic signing is a mechanical or algorithmic process should not succeed under \textit{Burrow-Giles}. Photography is mechanical; cryptography is algorithmic. Both employ technical processes that execute and record human choices. The presence of mechanical or mathematical operations does not defeat authorship when a work embodies the creator's original intellectual conceptions. Digital speech acts do precisely that: they combine expressive content, attribution, and volitional action into a unified work originating from the creator's own choices.

\subsection{Originality under \textit{Feist v.\ Rural} (1991)}\label{subsec:feist}

In \textit{Feist Publications, Inc. v. Rural Telephone Service Co.}, 499 U.S. 340 (1991), the Supreme Court held that Rural's white pages directory---an alphabetical listing of subscriber names and phone numbers---could not be copyrighted because it lacked the minimal creativity required for copyright protection.

The Court defined originality as requiring only that the work be ``independently created by the author'' and possess ``at least some minimal degree of creativity,'' emphasizing that ``the requisite level of creativity is extremely low; even a slight amount will suffice.''\footnote{\textit{Feist}, \textit{supra} note~\ref{fn:feist}, at 345.} Works need only possess ``some creative spark, `no matter how crude, humble or obvious' it might be.''\footnote{\textit{Id.}} Novelty is not required; independent creation plus minimal creativity suffices.\footnote{\textit{Id.} at 345--46.}

Digital speech acts easily satisfy this minimal standard because the copyrightable work—the digital speech act as a whole—embodies original expressive and volitional choices. The creator determines not only the substance of the expression, but also the communicative act being performed through it. A digital speech act may declare, promise, assert, direct, acknowledge, vote, transfer, or otherwise perform a social action. The creator further determines when the act will occur and, where multiple identities are available, with which cryptographic identity it will be associated. The resulting digital speech act—content, signature, and associated metadata—constitutes a fixed expression embodying those choices.

\textit{Feist} also clarifies the infringement standard: a plaintiff must prove ``ownership of a valid copyright'' and ``copying of constituent elements of the work that are original.''\footnote{\textit{Id.} at 361.} Because each digital speech act is cryptographically unique to the person who created it, both elements are readily established. The signature provides mathematical proof that only the holder of the corresponding private key could have created that particular authenticated expression.

That conclusion nevertheless raises a familiar concern in copyright theory. If the originality threshold is as low as \textit{Feist} suggests, extending protection to cryptographically authenticated digital speech acts may appear to risk overexpansion of copyright protection. Mark Lemley has demonstrated how copyright protection for works with minimal creativity can lead to strategic claiming over functional elements, creating anticommons problems that restrict follow-on creation and innovation.\footnote{\label{fn:lemley-property}Mark A. Lemley, \textit{Property, Intellectual Property, and Free Riding}, 83 Tex. L. Rev. 1031, 1059--65 (2005); Mark A. Lemley, \textit{The Economics of Improvement in Intellectual Property Law}, 75 Tex. L. Rev. 989 (1997).} Pamela Samuelson has cautioned against courts applying minimal creativity standards so permissively that formulaic or purely functional expressions receive protection.\footnote{\label{fn:samuelson-originality}Pamela Samuelson, \textit{The Originality Standard in Copyright Law}, 42 Am. J. Comp. L. 393 (1994).} We address these concerns in Part~IX.

\subsection{Fixation under 17 U.S.C. § 102(a)}

Digital speech acts satisfy the Copyright Act's fixation requirement under any defensible standard. The Act requires that works be ``fixed in any tangible medium of expression\ldots from which they can be perceived, reproduced, or otherwise communicated, either directly or with the aid of a machine or device.''\footnote{17 U.S.C. § 102(a).}

Digital speech acts are stored persistently on the creator's device---in solid-state storage, hard drives, or other non-volatile memory. They remain accessible indefinitely, can be retrieved and displayed at any time, and persist independently of RAM or other volatile storage. This clearly satisfies fixation under any reasonable interpretation of the statute.

The appropriate standard for fixation comes from \textit{Cartoon Network v. CSC Holdings}, which held that copies must persist long enough to be ``perceived, reproduced, or otherwise communicated'' for ``more than transitory duration.''\footnote{\textit{Cartoon Network}, \textit{supra} note~\ref{fn:cablevision}, at 127--30.} Digital speech acts easily satisfy this standard: they are stored persistently on devices, remain accessible indefinitely until deliberately deleted, and can be perceived and communicated at any time through the device interface. Their duration far exceeds the 1.2-second buffer copies that \textit{Cartoon Network} held insufficiently fixed.

Even under the more permissive---and heavily criticized---standard of \textit{MAI Systems v. Peak Computer}, digital speech acts would satisfy fixation. \textit{MAI} held that even temporary loading of software into RAM constitutes sufficient fixation.\footnote{\textit{MAI Systems}, \textit{supra} note~\ref{fn:mai}}This holding has been subject to substantial scholarly criticism. Pamela Samuelson argued that \textit{MAI}'s reasoning was doctrinally flawed, noting that the Copyright Act's definition of ``fixed'' requires works to be ``sufficiently permanent or stable to permit [them] to be perceived\ldots for a period of more than transitory duration,'' a requirement that RAM copying---which is lost when a computer powers off---arguably fails to meet.\footnote{Pamela Samuelson, \textit{The Copyright Grab}, Wired (Jan. 1996), https://archive.gyford.com/1997/wired-uk/2.01/features/copyright.html; see also Jessica Litman, \textit{The Exclusive Right to Read}, 13 Cardozo Arts \& Ent. L.J. 29, 34--37 (1994) (criticizing MAI's expansive fixation holding as inconsistent with the Copyright Act's text and purpose).} Jessica Litman similarly criticized \textit{MAI} for creating copyright liability in routine computer operations that Congress did not intend to regulate.\footnote{\textit{Id.}}

We do not rely on \textit{MAI}'s permissive standard. Digital speech acts satisfy the more demanding \textit{Cartoon Network} standard because they persist in non-volatile storage for indefinite duration, not merely in temporary RAM. If courts were to apply \textit{MAI}'s approach, digital speech acts would satisfy fixation \textit{a fortiori}---but the better view, supported by both statutory text and scholarly criticism, recognizes that true fixation requires meaningful persistence, which device-based storage of digital speech acts provides.

Whether courts apply \textit{Cartoon Network}'s ``more than transitory duration'' standard, \textit{MAI}'s permissive RAM-copying approach, or adopt an even more restrictive fixation doctrine, digital speech acts satisfy the requirement. They are embodied in tangible media (device storage), persist for more than transitory duration (indefinitely until deleted), and can be perceived and communicated at any time through the device interface.

\subsection{Copyright and Architecture}

Existing copyright doctrine provides a substantial basis for recognizing digital speech acts as copyrightable works. Under \textit{Burrow-Giles}, they embody personal creative choices despite the use of algorithmic processes. Under \textit{Feist}, those choices easily satisfy the minimal originality threshold. And because digital speech acts are persistently stored on personal devices, they satisfy the Copyright Act's fixation requirement.

If that analysis is accepted, an important implication follows. The same body of copyright law that helped enable platform dominance also contains doctrines that support distributed ownership of digital expressions. The difference is not a difference in law but a difference in architecture. Those architectural choices determine which copyright doctrines become practically significant and who is positioned to exercise the rights copyright confers.

If digital speech acts qualify as original works of authorship, the individual who creates them not only retains possession of the underlying content and data, but is also the author for purposes of copyright law and holds the exclusive rights attached to authorship. The significance of this conclusion extends beyond ownership itself. It exposes a deeper tension between two competing conceptions of participation in digital life: one grounded in persons exercising legal agency through rights they possess, and another grounded in participation mediated through permissions granted by intermediary institutions.        
\section{The User/Person Distinction}\label{sec:user-person}

This Article consistently distinguishes between \textit{users} and \textit{persons}. The distinction is not stylistic but analytical, reflecting how different architectural arrangements construct different legal subjects and enable different legal relationships.

\textbf{We use the term ``users''—with all the legal and economic implications that term carries—when discussing centralized platforms.} Platform participants are users: a discrete legal category created by software licensing, defined by license-bound access, adhesive agreements, asymmetric obligations, and possession-surrender.

\textbf{We use the term ``persons''—both natural persons (human beings) and juridical persons (corporations, organizations)—when discussing grassroots networks.} Grassroots participants are persons: owners of devices, possessors of private keys, creators of copyrighted expressions, and parties to voluntary digital social contracts.

The distinction tracks a fundamental architectural difference that produces altering legal subjects.

\subsection{``User'' as a Legal Category Created by Platform Architecture}

A \textit{user} is a discrete legal category that emerges from software licensing applied at platform scale. The term entered legal discourse through the software industry's characterization of people who access and use software as distinct from those who own it.\footnote{See Madison, \textit{supra} note~\ref{fn:madison}, at 280--85 (tracing emergence of ``user'' as distinct from ``owner'' through software licensing practice).} This distinction was crucial to circumventing copyright's first-sale doctrine: if transactions are characterized as licenses rather than sales, recipients are "users" bound by license terms rather than "owners" entitled to copyright exemptions.\footnote{See 17 U.S.C. § 109(a) (first sale doctrine applies to "owner of a particular copy"); \textit{MAI Systems}, \textit{supra} note~\ref{fn:mai} (software recipients are "licensees" not "owners" and thus not entitled to § 117 exemptions).}

The "user" as a legal subject is defined by four characteristics, each carrying specific legal and economic implications:

\begin{enumerate}[label=(\arabic*),leftmargin=2em]

\item \textbf{License-bound status}: Users access software and platforms through revocable licenses rather than ownership interests. Because access is licensed rather than sold, users possess no first-sale rights, no ability to transfer access, and little protection against termination. This arrangement also permits platforms to extract ongoing rents for access that, under a sale model, would require only a one-time transaction.\footnote{See \textit{ProCD}, \textit{supra} note~\ref{fn:procd},
at 1450--52 (enforcing license prohibiting commercial use despite purchaser buying retail copy).}

\item \textbf{Adhesive agreement}: Users participate subject to non-negotiable terms drafted unilaterally by platform operators. Because access is conditioned upon acceptance of the entire agreement, users cannot meaningfully negotiate, modify, or reject particular provisions while retaining participation. The result is a contractual environment in which platforms may impose expansive content and data licenses, liability limitations, and other terms on a take-it-or-leave-it basis.\footnote{See Nancy S. Kim, \textit{Wrap Contracts: Foundations and Ramifications} 47--73 (2013) (analyzing adhesive nature of software licenses).}

\item \textbf{Asymmetric obligations}: Users grant extensive rights over their content, data, and activity while receiving only revocable access to platform services. Platforms routinely obtain transferable, sublicensable, royalty-free rights in user-generated content, whereas users receive access that may be restricted or terminated at the platform's discretion. This asymmetry enables platforms to monetize user activity through advertising, data analytics, training datasets, and derivative services without corresponding ownership interests for participants.\footnote{See infra Part~\ref{sec:platform-regime}.E (examining contemporary platform Terms of Service).}

\item \textbf{Surrender of possession}: Users upload content to provider-controlled infrastructure, relinquishing possession while retaining nominal copyright ownership. As a consequence, platforms control access, distribution, modification, and preservation of content regardless of who legally owns the underlying expression. Possession therefore becomes a source of practical authority, allowing platforms to mediate how users access and derive value from their own content.\footnote{See Cohen, \textit{supra} note~\ref{fn:cohen-truth}, at 37--65 (platforms obtain control through possession regardless of nominal ownership).}

\end{enumerate}

Platforms operationalize this legal category at global scale. Meta's 3.58 billion Family Daily Active People (DAP) as of December 2025,\footnote{\textit{Meta 2025 Form 10-K}, supra note~\ref{fn:meta2025}} YouTube's more than 2 billion monthly logged-in users,\footnote{YouTube, Press Page, https://www.youtube.com/about/press/ (last visited May 15, 2026).} and X's hundreds of millions of users all occupy the "user" subject position: license-bound, adhesively-contracted, asymmetrically-obligated, possession-surrendering participants in platform-controlled infrastructure. The modern user is therefore not a natural legal category but the product of software licensing doctrines that transformed owners into licensees and were subsequently generalized across digital platforms.\footnote{For a detailed genealogy of this development, see infra note~\ref{fn:prehistory}; \textit{MAI Systems}, \textit{supra} note~\ref{fn:mai}, at 518 n.5; \textit{ProCD}, \textit{supra} note~\ref{fn:procd}, at 1450--52.}

\subsection{Personhood in Grassroots Architectures}

In contrast, grassroots social networking has no architectural position for an entity to construct "users" as a legal category. With no intermediary to draft Terms of Service, no servers to which content must be uploaded, and no central authority conditioning access on license grants, the relevant legal subject is the \textit{person}—natural or juridical—who participates through devices they own and expressions they create.

In U.S. law, "person" includes both natural persons (human beings) and juridical persons (corporations, companies, associations, partnerships, and other legal entities).\footnote{1 U.S.C. § 1 ("the word 'person' ... include[s] corporations, companies, associations, firms, partnerships, societies, and joint stock companies, as well as individuals").} \textbf{Both natural and juridical persons can participate in grassroots networks.} A corporation could operate through devices it owns, authenticate through private keys it possesses, and create digital speech acts it copyrights—just as a human being could. The legal category is "person" not "user" because no platform exists to construct the user-subject through license-mediation.

A \textit{person} in grassroots architecture is defined by four characteristics that mirror and invert the user-subject position:

\begin{enumerate}[label=(\arabic*),leftmargin=2em]

\item \textbf{Ownership not license}: Persons own the devices on which they operate (smartphones for natural persons, servers for juridical persons), the private keys through which they authenticate, and the copyright in expressions they create. Because participation rests upon ownership rather than licensed access, persons retain full rights of transfer, exclusion, and control. They may also derive value from their own property and expressions without requiring intermediary permission or surrendering rents to platform operators.

\item \textbf{Voluntary agreement not adhesion}: Persons participate through digital social contracts (voluntary agreements among participants, specified and enforced through code).\footnote{Cardelli et al., \textit{supra} note~\ref{note:cardelli}.} Participation remains optional, the governing rules are visible, and exit carries no penalty beyond loss of access to that particular network. Because no intermediary controls participation, no entity can impose extractive contractual terms as a condition of entry.

\item \textbf{Symmetric obligations}: Persons entering digital social contracts accept the same rules as all other participants. No participant occupies a privileged legal position through which rights are accumulated while obligations are externalized. The result is a more symmetrical relationship among participants, in which value cannot be systematically extracted from others' contributions without providing corresponding value in return.

\item \textbf{Retained possession}: Persons store their own data on their own devices. Nothing is uploaded to third-party servers and nothing is surrendered to another entity's possession. As a consequence, possession and ownership remain aligned: the person who owns the expression also possesses it. Without intermediary possession, no platform can unilaterally monetize, manipulate, restrict, or control access to that content.

\end{enumerate}

\subsection{Ownership, Possession, and Control}

The distinction between users and persons is not  terminological. It identifies two different legal subject positions produced by two different architectures. 

Much contemporary scholarship seeks to improve the position of users within platform systems through stronger privacy protections, portability rights, or limitations on contractual overreach. Such proposals generally accept the continued existence of platform-mediated relationships and focus on modifying the terms under which users participate. The present analysis instead asks what legal consequences follow when the architectural conditions that produce the user-subject are absent altogether.

Grassroots architectures do not merely offer users better terms. They
eliminate the architectural conditions through which users are
constructed as a distinct legal category. Where platforms
position individuals as licensees operating through accounts,
grassroots systems position participants as persons operating
through devices they own, keys they control, and expressions
they author.

The distinction becomes particularly significant when copyright ownership is considered. Copyright law vests rights in authors and owners, not in users as such. If digital speech acts qualify as copyrightable expression, the question is not whether a platform user possesses rights granted by a service provider, but whether a person possesses rights arising directly from authorship. The answer determines where ownership resides, who exercises control over expression, and whether copyright functions as a mechanism of dependence or autonomy. Copyright ownership thus becomes more than a private economic entitlement: it becomes a legal mechanism through which persons may exercise control over their digital expressions and the means through which those expressions are attributed, authenticated, and shared.        
\section{Copyright Ownership and Digital Sovereignty}\label{sec:ownership-sovereignty}

Part~\ref{sec:intro} introduced digital sovereignty as the ability to conduct one's digital life free of third-party control, surveillance, manipulation, and rent-seeking. Cryptographic possession can provide technical control over digital expressions, but digital sovereignty ultimately depends upon whether those expressions are also recognized as legally owned. Copyright ownership supplies that legal foundation, complementing the cryptographic possession inherent in device-based architectures.

The argument is not merely that copyright protection for digital speech acts would be beneficial. The claim is structural: grassroots architectures can provide cryptographic security and technical control, but without copyright recognition, people cannot enforce ownership claims through legal institutions or prevent unauthorized commercial exploitation. Copyright transforms technical possession of digital expressions into legally enforceable ownership, enabling digital sovereignty.

\subsection{Dual Ownership: Legal Copyright and Cryptographic Possession}

Digital speech acts implicate two complementary forms of authority, neither alone sufficient but together constitutive of digital sovereignty: cryptographic possession and legal ownership.

\textbf{Cryptographic possession.} When a person creates a digital speech act, they exercise factual control through private key cryptography. Only the person possessing the private key can produce valid signatures. This creates a form of technical possession enforced by mathematical impossibility of forgery—no intermediary can create, modify, or repudiate a digital speech act without the private key. The expression is stored on the person's own device, under their exclusive physical control.

Cryptographic possession is factual control over a digital object, not legal possession in property law's doctrinal sense.\footnote{On the distinction between factual control and legal possession, see Carol Rose, \textit{Possession as the Origin of Property}, 52 U. Chi. L. Rev. 73 (1985); Henry E. Smith, \textit{Property as the Law of Things}, 125 Harv. L. Rev. 1691 (2012).} A person may have exclusive technical control over data on their device while lacking legal rights to prevent others from copying, distributing, or commercially exploiting that data if they can obtain it through other means. Cryptographic possession provides control and security, but not legal ownership.

\textbf{Legal copyright.} Copyright provides legal ownership but, standing alone in platform architectures, proves insufficient for digital sovereignty. A person may retain copyright in content they create while being structurally compelled to license that copyright to platforms as the condition of participation. Legal ownership without control over infrastructure creates the dependency relationship that enables platform extraction.

If digital speech acts qualify as copyrightable works under Part~\ref{sec:core}'s analysis, that legal ownership takes the form of the exclusive rights enumerated in Section 106: reproduction, preparation of derivative works, distribution, public performance, and public display.\footnote{17 U.S.C. \S~106.} These rights are enforceable through legal institutions against third parties who copy, distribute, or commercially exploit the work without authorization.

\textbf{Dual ownership in grassroots architectures.} When people create and store digital speech acts on their own devices using their own private keys, and copyright law recognizes them as authors of those expressions, cryptographic possession and legal ownership converge. The person has both factual control (technical possession via private key) and legal rights (copyright ownership via authorship). No intermediary controls the infrastructure; no intermediary can compel licensing; no license must be granted as condition of participation.

This dual ownership—technical and legal—is what distinguishes grassroots from platform architectures. Grassroots architectures unify possession and ownership: the person possesses the content (stored on personal device) and owns it legally (copyright via authorship). The convergence of cryptographic possession and legal ownership closes the possession-ownership gap identified earlier and provides the ownership structure on which digital sovereignty depends.

\subsection{From Ownership to Sovereignty}

Dual ownership establishes the conditions for digital sovereignty because it reunites legal ownership and practical control within the same actor. Where possession and ownership converge in the same person, no intermediary can exercise authority solely by virtue of possessing the underlying data, expressions, or infrastructure. Device-based architecture provides the technical foundation: individuals control the infrastructure through which they communicate and authenticate. Copyright provides the legal foundation: individuals own the expressions they create and may enforce that ownership against unauthorized appropriation. Together, these forms of control align possession and ownership in the same person.

Digital sovereignty therefore follows not from ownership alone but from the distribution of ownership. The significance of copyright recognition is thus not merely economic. Copyright transforms digital speech acts from expressions that can be technically controlled into expressions that can also be legally owned. Combined with device-based architecture, that ownership structure disperses control across participants and removes the centralized points of authority through which surveillance, manipulation, and rent-seeking become possible.

If digital speech acts do not qualify as copyrightable works, grassroots architectures may provide cryptographic security but not legal ownership. Participants would possess their expressions technically without possessing enforceable rights against appropriation or exploitation. Copyright recognition completes the ownership structure. By aligning legal ownership with cryptographic possession and distributing both across participants, it provides a necessary foundation for digital sovereignty.

\section{Provenance, Accountability, and Legal Culpability}\label{sec:provenance}

Copyright ownership does not by itself resolve questions of responsibility. A system in which individuals own their expressions but cannot determine who created, modified, or redistributed them remains vulnerable to manipulation, misinformation, and attribution disputes. Digital sovereignty concerns not only the distribution of ownership but also the distribution of accountability.

The same cryptographic architecture that aligns possession and ownership also creates verifiable provenance. Every digital speech act is signed by its creator, and every forwarding event generates a new signed expression that incorporates the prior one. Provenance therefore records not only authorship but also the history of dissemination. It becomes possible to identify who created content, who redistributed it, and in what sequence.

These provenance chains raise a distinct set of legal questions. How does copyright operate when digital speech acts are forwarded through successive participants? Does the original author's copyright persist through subsequent dissemination? What responsibilities attach to those who redistribute content? How are provenance records authenticated as evidence? And how do questions of jurisdiction arise when dissemination crosses national borders?

\subsection{Cryptographic Provenance Chains}

A digital speech act is either an original expression---a person deliberately signs content with their private key---or a forwarding expression---a person signs a new digital speech act whose content consists of a previously received digital speech act. In the forwarding case, the forwarded speech act is embedded within the new one, including the original creator's signature.

When content is forwarded multiple times, each forwarding creates a new digital speech act that embeds the preceding one. The result is a chain: the outermost speech act is signed by the most recent forwarder and contains, as content, the speech act of the previous forwarder, which in turn contains the speech act before it, continuing back to the original creator's speech act. Each link in the chain is cryptographically signed by the person who created it, and the chain as a whole constitutes a verifiable record of the content's provenance---who created it, who forwarded it, and in what order.\footnote{\textit{See} Shapiro, \textit{supra} note~\ref{note:gsn} (``utterances that have been forwarded multiple times are provided with their entire provenance'').}

The provenance chain is not external to the expression but part of the expression itself. Records of creation, forwarding, and dissemination therefore travel with the content wherever it goes, creating a persistent and independently verifiable history of the expression's origin and transmission. Because that history is embedded within the digital speech act, it does not depend upon intermediary recordkeeping for its preservation or verification.

\subsection{Copyright Preserved at Each Step}

From a copyright perspective, forwarding is additive rather than substitutive. The original digital speech act remains fixed and intact within the forwarding speech act, together with the cryptographic signature that identifies its author. The forwarder creates a new digital speech act incorporating the earlier one, but does not displace or alter the original work. As a result, the chain contains multiple copyrightable expressions layered together, each retaining its own authorship and legal status within the provenance chain.

This structure is analogous to republishing a book with a new foreword. The foreword may itself be copyrightable, but its creation does not diminish or replace the copyright in the underlying book. Likewise, each forwarding digital speech act may constitute a new copyrightable expression, while the embedded digital speech acts retain their independent copyrights. The cryptographic structure makes this relationship explicit and verifiable: the original creator's signed expression remains embedded within every subsequent forwarding, regardless of how many times the content is redistributed.

\subsection{Accountability and Legal Liability}\label{subsec:accountability}

Each digital speech act is cryptographically signed by its creator, establishing verifiable authorship and deliberate choice to create or forward content. This cryptographic accountability has complex legal implications that differ substantially from platform architectures—in some ways increasing individual responsibility, in others raising unresolved questions about the application of existing intermediary liability frameworks.

\subsubsection{Forwarder Liability: Unresolved Questions}

Grassroots architectures eliminate platform intermediaries, raising the question: what liability do individual forwarders face? The answer is doctrinally complex and depends on how courts apply existing frameworks to peer-to-peer forwarding.

\textbf{Section 230 and grassroots participants.} Section 230's immunity extends to ``users'' of interactive computer services, not only to service providers.\footnote{47 U.S.C. \S~230(c)(1) (``No provider \textit{or user} of an interactive computer service....'') (emphasis added).} Courts have applied this protection to individuals who republish content created by others online.\footnote{See \textit{Batzel v. Smith}, 333 F.3d 1018, 1030--32 (9th Cir. 2003) (operator of email listserv qualified for \S~230 immunity); \textit{Barrett v. Rosenthal}, 40 Cal. 4th 33, 41--61 (2006) (individual who reposted allegedly defamatory messages to Usenet newsgroup entitled to \S~230 immunity).} In \textit{Barrett}, the California Supreme Court held that an individual who redistributed allegedly defamatory messages qualified as a protected ``user'' under Section 230.\footnote{\textit{Barrett}, 40 Cal. 4th at 51--54.}

Whether a person participating in a grassroots provenance network would be treated as a "user of an interactive computer service" for purposes of Section 230 remains unresolved. Section 230's text extends protection to "users" of interactive computer services, and courts have applied that protection to individuals who republish third-party content online. On one view, a person who forwards a digital speech act may fall within that category in much the same way that the defendant in \textit{Barrett} was treated as a user of Usenet. On another view, grassroots architectures differ in legally significant respects. Because grassroots participants act simultaneously as creators, distributors, and recipients rather than through a centralized intermediary, existing Section 230 doctrine provides only an imperfect analogue. Section 230 does not protect actors who materially contribute to allegedly unlawful content or who become "information content providers" through the creation or development of the challenged expression.\footnote{See \textit{Fair Hous. Council of San Fernando Valley v. Roommates.com, LLC}, 521 F.3d 1157, 1168--69 (9th Cir. 2008) (en banc) (no immunity when website materially contributed to alleged illegality).} A participant who adds commentary, selectively redistributes content to amplify harmful material, or knowingly propagates falsehoods may fall outside the statute's protection.

The cryptographic provenance structure introduces an additional uncertainty. Every forwarding event is affirmatively signed, creating a verifiable record that a particular person chose to redistribute the content. Courts may conclude that such deliberate, attributable forwarding remains protected republication activity under Section 230. Alternatively, they may view cryptographically authenticated forwarding as evidence of a level of editorial participation sufficient to place the actor outside the statute's intended protections. Existing doctrine provides no definitive answer.

Accordingly, the application of Section 230 to grassroots forwarding remains unresolved. The provenance chain strengthens attribution and accountability, but whether those same features preserve or diminish statutory immunity is ultimately a question for future courts.

\textbf{DMCA safe harbors and grassroots participation.} The DMCA's Section 512 safe harbors protect certain ``service providers'' from copyright infringement liability arising from the activities of others, provided they satisfy specified statutory requirements.\footnote{17 U.S.C. \S~512.} The statute defines a service provider broadly to include ``an entity offering the transmission, routing, or providing of connections for digital online communications.''\footnote{17 U.S.C. \S~512(k)(1)(B).} Whether participants in a grassroots architecture fall within this definition is uncertain.

Courts have applied Section 512 to a variety of intermediaries, but the paradigm cases involve entities operating communications infrastructure or online services for the benefit of others.\footnote{See, e.g., \textit{Ellison v. Robertson}, 357 F.3d 1072, 1076--80 (9th Cir. 2004) (applying \S~512 to Internet access provider).} A person who receives and forwards digital speech acts through their own device occupies a different position. Such a participant is not obviously providing a communications service to the public; rather, they are engaging in expressive activity through the creation and forwarding of digital speech acts.

Accordingly, it is unclear whether Section 512's safe harbors would apply to grassroots participants. Courts may conclude that forwarding activity falls outside the category of protected intermediary conduct, leaving questions of contributory or vicarious copyright infringement to be resolved under traditional copyright doctrines.

\textbf{Defamation and the republication rule.} The traditional tort rule holds that ``[o]ne who repeats or otherwise republishes defamatory matter is subject to liability as if he had originally published it.''\footnote{Restatement (Second) of Torts \S~577(1) (1977).} This suggests that a person who forwards a defamatory digital speech act may incur liability through the act of republication itself. However, the single-publication rule complicates that conclusion.

The single-publication rule provides that ``the publication of a writing... gives rise to only one claim for relief for libel.''\footnote{\textit{Firth v. State}, 98 N.Y.2d 365, 371 (2002); see also Restatement (Second) of Torts \S~577A (1977).} The rule was developed to prevent mass-media defendants from facing separate suits for every copy of a newspaper or book distributed.\footnote{See \textit{Keeton v. Hustler Magazine, Inc.}, 465 U.S. 770, 777 n.8 (1984) (describing the single-publication rule).} Courts later extended the doctrine to online publications, treating the posting of content on a website as a single publication regardless of the number of subsequent views.\footnote{See \textit{Firth}, 98 N.Y.2d at 371--72 (single-publication rule applies to online postings).}

Whether the forwarding of a digital speech act constitutes a new publication or merely extends an existing one is uncertain. On one view, forwarding resembles the continued circulation of an already published work and therefore falls within the logic of the single-publication rule. On another view, each forwarding event intentionally transmits the expression to a new audience and therefore constitutes a distinct republication. Courts have long distinguished between passive availability and active redistribution.\footnote{See \textit{Hoffman v. Capital Cities/ABC, Inc.}, 255 F.3d 1180, 1185 (9th Cir. 2001) (passive continued availability of online content does not create new publication; republication may occur when material is directed to a new audience).} Deliberate peer-to-peer forwarding appears more analogous to the latter, but existing doctrine provides no definitive answer.

Existing defamation law was developed in the context of publishers, broadcasters, and platform-hosted communications rather than cryptographically attributable provenance chains. As a result, courts could conclude that forwarding creates independent republication liability, or they could treat forwarding as part of a broader continuing publication. The provenance chain does not determine the legal outcome, but it does provide unusually clear evidence regarding the conduct at issue. Each forwarding event is independently attributable to a particular person and can be verified cryptographically. The legal significance of that attribution remains uncertain, but the evidentiary record is substantially clearer than in conventional online environments.

\textbf{Copyright and contributory infringement.} A person who forwards infringing content may face contributory copyright infringement liability. The traditional test requires: (1) knowledge of the infringing activity, and (2) material contribution to the infringement.\footnote{\textit{Gershwin Publ'g Corp. v. Columbia Artists Mgmt., Inc.}, 443 F.2d 1159, 1162 (2d Cir. 1971).} A participant who knowingly forwards infringing content contributes to its continued distribution, potentially satisfying both elements depending on the circumstances of the forwarding and the extent of the contribution.

Separate from contributory infringement is inducement liability. In \textit{MGM Studios Inc. v. Grokster, Ltd.}, the Supreme Court held that ``one who distributes a device with the object of promoting its use to infringe copyright... is liable for the resulting acts of infringement by third parties.''\footnote{545 U.S. 913, 919 (2005).} Inducement requires proof of purposeful, culpable expression and conduct directed toward encouraging infringement.\footnote{\textit{Id.} at 936--37.} Participants who merely forward content are unlikely to incur inducement liability absent evidence that they encouraged or promoted infringement, but the doctrine remains available in more egregious cases.

The provenance chain does not determine whether liability exists, but it does provide evidence relevant to the inquiry. Because each forwarding event is independently signed, the chain identifies the sequence of participants through which content was distributed. This may assist in establishing questions of knowledge, participation, and causation that often prove difficult in conventional online environments. At the same time, liability remains dependent on the specific facts of each case. A person who forwards content without knowledge of infringement may lack the requisite scienter, while a person whose forwarding plays only a minimal role in dissemination may not satisfy traditional standards of material contribution.

\subsubsection{Accountability and Chilling Effects}

The foregoing discussion reveals a tension inherent in attributable communication systems. The same provenance mechanisms that provide transparency and accountability may also increase perceived legal risk. If participants believe that forwarding a digital speech act could expose them to defamation, copyright, or other claims, some may choose not to participate in forwarding at all.

Existing defenses remain available, including truth in defamation actions, fair use in copyright disputes, and the absence of knowledge where liability doctrines require scienter. Moreover, provenance chains may assist not only plaintiffs but also defendants by providing evidence regarding the origin of content, the scope of dissemination, and the participant's role within the chain.

Nonetheless, uncertainty itself can produce caution. Existing liability doctrines developed in environments characterized by publishers, intermediaries, and platform-hosted communications. Their application to cryptographically attributable provenance chains remains unsettled. Until courts clarify how these doctrines apply, some participants may perceive forwarding as carrying greater legal risk than comparable conduct on platform-mediated networks.

Whether this increased accountability ultimately discourages participation or instead promotes more deliberate and responsible dissemination practices is an empirical question beyond the scope of this Article. The important point for present purposes is that provenance changes not only the evidentiary landscape but also the incentives surrounding redistribution.

\subsubsection{Accountability Through Provenance}

The preceding discussion does not yield definitive answers regarding liability. Existing doctrines governing intermediary immunity, republication, and contributory infringement developed in legal environments that did not contemplate cryptographically attributable provenance chains. Whether courts extend, modify, or limit those doctrines in the context of digital speech acts remains uncertain.

What is less uncertain is the effect of provenance on accountability. Provenance chains provide a transparent record of authorship and dissemination that is largely absent from contemporary platform-mediated communications.

This transparency does not itself determine liability. A provenance chain cannot establish whether a statement is defamatory, whether a participant possessed the requisite knowledge for contributory infringement, or whether a particular immunity applies. Those questions remain matters of substantive law. Provenance instead affects the evidentiary foundation upon which those questions are resolved.

In this respect, grassroots architectures invert a defining characteristic of contemporary platforms. Platform systems frequently rely upon centralized moderation, opaque recommendation mechanisms, and institutional records to determine responsibility. Provenance-based architectures distribute expressive authority among participants while simultaneously preserving a verifiable record of their actions. The result is not immunity from legal responsibility but a framework in which responsibility can be traced with greater precision.

Whether this shift ultimately advances democratic values, free expression, or effective governance is a broader normative question addressed in subsequent Parts. For present purposes, the relevant point is narrower: digital speech acts transform accountability from an inference drawn from platform records into a property embedded within the architecture itself.

\subsection{Evidentiary Authentication of Provenance Chains}

The accountability value of provenance depends upon more than technical verifiability. Provenance chains may record authorship, forwarding, and dissemination with cryptographic precision, but those records must still be admissible in legal proceedings before they can support legal rights or obligations. The question therefore becomes how a provenance chain proves its authenticity to a court. This issue is central to any litigation over digital speech acts and requires examination of authentication requirements under the Federal Rules of Evidence.

\subsubsection{The Authentication Requirement}

To be admissible, evidence must be authenticated: the proponent must ``produce evidence sufficient to support a finding that the item is what the proponent claims it is.''\footnote{Fed. R. Evid. 901(a).} For a digital speech act, authentication requires establishing that: (1) the content was created or forwarded by the person whose signature appears on it, and (2) the content has not been altered since signing.

Traditional authentication of digital evidence involves witness testimony (``I sent this email''), distinctive characteristics (``this appears to be from the defendant's email account''), or chain of custody evidence.\footnote{See Fed. R. Evid. 901(b)(1), (4), (9).} Digital speech acts present a different model of authentication: cryptographic proof that a signature is mathematically valid for a given public key.

\subsubsection{Self-Authenticating Digital Signatures}

Rule 902(14), added in 2017, permits self-authentication of electronically stored information authenticated through a reliable process of digital identification. The Advisory Committee Notes specifically contemplate authentication through hash values and similar cryptographic techniques.

A digital speech act's cryptographic signature operates similarly: it is a mathematical function of the content and private key that can be verified using the corresponding public key. If verification succeeds, the signature mathematically proves the content has not been altered since signing and was signed by whoever holds the private key corresponding to the public key. This process—cryptographic verification—is a ``process of digital identification'' under Rule 902(14).

\subsubsection{Authentication Procedure}

Authentication under Rule 902(14) would require several related showings. First, the proponent would present the digital speech act itself, including the content, signature, timestamp, and public-key identifier. Second, the proponent would demonstrate that cryptographic verification of the content, signature, and public key produces a valid result, establishing that the signed content has not been altered and that it was signed using the corresponding private key.

Third, the proponent would need to link the public key to the person alleged to have created or forwarded the digital speech act. This remains the critical evidentiary step. Cryptographic verification can establish that a particular key signed a particular message, but additional evidence is ordinarily required to establish who controlled that key. Such evidence may include testimony, prior use of the public key in communications, public registration of the key, admissions by the purported signer, or circumstantial evidence connecting the content to a particular individual.

Finally, Rule 902(14) contemplates certification of the digital-identification process and its reliability. Once these elements are established, the digital speech act may be authenticated through cryptographic verification without requiring the traditional forms of extrinsic authentication commonly associated with electronic evidence.

\subsubsection{Provenance Chain as Business Record}

Alternatively, a provenance chain may be admissible through the business-records exception under Rule 803(6) where records of digital speech acts are systematically maintained in the regular course of an organization's activities.\footnote{Fed. R. Evid. 803(6).} The cryptographic signatures and timestamps would be records ``kept in the course of a regularly conducted activity'' if the grassroots application automatically records them. This approach requires testimony from a custodian or qualified witness that the records are maintained systematically and reliably.

However, the business record approach faces limitations: the hearsay exception addresses the record-keeping process, not the truth of the content. A business record establishes that \textit{a signature was recorded}, not that the signature is \textit{valid}. Cryptographic authentication under Rule 902(14) is stronger: it proves mathematical validity of the signature, not merely that someone recorded it.

\subsubsection{Challenges and Limitations}

Cryptographic authentication significantly simplifies evidentiary questions, but it does not eliminate them. A valid digital signature proves that a particular private key signed a particular expression and that the signed content has not been altered. It does not, by itself, resolve every factual question relevant to litigation.

\textbf{Linking public keys to persons.} The most significant limitation is attribution. Signature verification establishes that a digital speech act was signed by the private key corresponding to a particular public key, but a court must still determine who controlled that key. If a defendant denies ownership, the proponent must establish the connection through testimony, prior communications, public registration records, admissions, or other circumstantial evidence. This challenge resembles traditional authentication disputes over email accounts or social media profiles: proving that a communication originated from a particular account does not automatically prove who controlled the account at the relevant time.\footnote{See, e.g., \textit{United States v. Siddiqui}, 235 F.3d 1318, 1322--23 (11th Cir. 2000) (authentication of email requires proof of authorship).}

\textbf{Key compromise.} A defendant may contend that a private key was lost, stolen, or otherwise compromised. Cryptographic verification cannot independently disprove such a claim; it establishes only that the key was used. As with traditional allegations of forgery or unauthorized account access, the factfinder must evaluate the credibility of the claim in light of surrounding evidence, including the circumstances of the communication and the defendant's prior use of the key.

\textbf{Timestamp reliability.} Digital speech acts may include timestamps, but the signature proves only that the timestamp formed part of the signed expression. It does not necessarily establish that the recorded time is accurate. If the device clock was incorrect or manipulated, additional evidence may be necessary to verify when the communication actually occurred. Corroborating evidence might include receipt by other participants, contemporaneous records, or references within the content itself.

\textbf{Expert testimony and technical understanding.} Although Rules 901 and 902 permit authentication through digital identification processes, courts and juries may still require assistance understanding how cryptographic verification operates. Expert testimony may therefore be necessary to explain the underlying technology, the significance of a valid signature, and the limits of the conclusions that can properly be drawn from it.\footnote{See, e.g., \textit{In re Vee Vinhnee}, 336 B.R. 437, 444--45 (9th Cir. BAP 2005); \textit{United States v. Simpson}, 152 F.3d 1241, 1250 (10th Cir. 1998).}

These limitations define the scope of what cryptographic authentication establishes. A provenance chain can provide unusually strong proof that a particular expression was signed by a particular key and remained unchanged thereafter. Questions of key ownership, authorization, and surrounding context remain matters for conventional evidentiary proof.

\subsubsection{Comparative Advantage Over Platform Evidence}

Despite these limitations, provenance chains offer significant evidentiary advantages over conventional platform-hosted communications.

Authentication of platform content typically depends upon records maintained by the platform itself. A party seeking to authenticate a social-media post, message, or account activity often relies on business records, account logs, metadata, or testimony supplied by the platform operator. Although such evidence may be sufficient under traditional evidentiary standards, its reliability ultimately depends upon the accuracy of the platform's record-keeping systems and the availability of the platform to provide the relevant information.

Digital speech acts shift much of this authentication burden from institutional record keeping to cryptographic verification. A valid digital signature directly demonstrates that a particular public key signed a particular expression and that the signed content has not been altered thereafter. Authentication therefore depends less upon the records of a third-party intermediary and more upon mathematical verification of the communication itself.

This distinction does not eliminate ordinary evidentiary disputes. Questions of key ownership, authorization, compromise, and attribution remain subject to proof through conventional means. But once the public key is linked to a person, the provenance chain provides unusually strong evidence of integrity and transmission history without requiring reliance upon a platform's internal records.

The evidentiary advantages become particularly significant when communications cross institutional or national boundaries. Authentication of platform-hosted content may require cooperation from platform operators, access to internal records, or compliance with foreign legal processes. By contrast, the cryptographic proof embedded within a provenance chain travels with the communication itself. The content, signatures, and verification data remain available regardless of where the communication originated or which jurisdictions later become involved in the dispute.

\subsubsection{Provenance, Evidence, and Accountability}

Existing evidentiary doctrine is well equipped to accommodate provenance chains. Rule 902(14)'s self-authentication framework for digitally identified data provides a natural mechanism for authenticating cryptographically signed digital speech acts, while traditional evidentiary tools remain available to establish the connection between public keys and particular persons. Challenges persist---including key attribution, claims of key compromise, timestamp accuracy, and the need to explain cryptographic verification to factfinders---but these are familiar evidentiary questions rather than novel barriers to admissibility.

The significance of provenance chains lies not merely in their technical reliability but in their evidentiary structure. Unlike platform-hosted content, whose authentication often depends on institutional records and third-party cooperation, digital speech acts carry their own proof of integrity and provenance. Authentication therefore rests on cryptographic verification rather than the accuracy of a platform's record-keeping systems.

This evidentiary architecture reinforces the broader accountability framework discussed in this Part. Provenance chains provide a verifiable record of authorship and dissemination, making it possible to identify who created content, who forwarded it, and in what sequence. Existing evidentiary rules can accommodate that record, enabling provenance information to be presented in court when disputes arise over ownership, attribution, or legal responsibility.

\subsection{International and Jurisdictional Complexity}

The grassroots architecture's elimination of platform intermediaries does not eliminate jurisdictional questions; it redistributes them. Digital speech acts may originate in one country, be forwarded through participants in another, and cause alleged harm in a third. The resulting provenance chain creates a transparent record of authorship and dissemination, but it also raises difficult questions of personal jurisdiction, choice of law, and cross-border enforcement.

\subsubsection{The Limits of a U.S.-Centric Analysis}

The analysis throughout this Part is primarily grounded in United States law. Discussions of Section 230, contributory copyright infringement, evidentiary authentication, personal jurisdiction, and defamation doctrine necessarily rely on U.S. statutes and precedent because those are the legal authorities governing the questions examined in this Article.

The emphasis on U.S. law is also significant because many contemporary debates about digital communication have been shaped by American doctrines governing intermediary liability and platform governance. Section 230 of the Communications Decency Act, for example, limits the circumstances under which online platforms and users may be treated as publishers of third-party content, while the Digital Millennium Copyright Act establishes safe harbors for qualifying service providers that facilitate online communications.\footnote{47 U.S.C. \S~230(c)(1); 17 U.S.C. \S~512.} These frameworks have profoundly influenced the legal architecture of platform-mediated communication in the United States and provide an important baseline against which grassroots architectures can be evaluated.

The underlying technological architecture is jurisdiction-neutral. A digital speech act carries the same cryptographic signature, provenance chain, and forwarding history whether created in New York, London, Singapore, or Berlin. What varies is the legal significance assigned to those facts.

The distinction is important. Provenance chains do not determine liability; they determine attribution. They establish who created a digital speech act, who forwarded it, and when those actions occurred. Whether those facts give rise to defamation liability, copyright liability, regulatory exposure, or no liability at all depends on the substantive law applied by the relevant jurisdiction.

For a United States audience, the most immediate questions concern how existing American doctrines would treat provenance-based communication systems. That inquiry provides a useful starting point because many of the legal questions discussed above—authentication, authorship, intermediary liability, and personal jurisdiction—have mature doctrinal frameworks in U.S. law. But the same provenance evidence could produce materially different outcomes under foreign legal systems that place different weight on speech rights, reputation interests, intermediary responsibility, or regulatory control.

Accordingly, the analysis offered here should be understood as jurisdictionally bounded. The provenance mechanisms are global; the legal consequences are local.

\subsubsection{Cross-Border Speech and Personal Jurisdiction}

When a person in Country A creates a digital speech act and someone in Country B forwards it to someone in Country C, which country's law applies? Which courts have jurisdiction over the creator and forwarders?

\textbf{Personal jurisdiction over digital speech.} U.S. courts exercise personal jurisdiction over out-of-state defendants based on ``minimum contacts'' with the forum such that the exercise of jurisdiction ``does not offend traditional notions of fair play and substantial justice.''\footnote{\textit{International Shoe Co. v. Washington}, 326 U.S. 310, 316 (1945).} For intentional torts like defamation, the ``effects test'' permits jurisdiction where a defendant expressly aimed conduct at the forum state and the brunt of the harm was suffered there.\footnote{\textit{Calder v. Jones}, 465 U.S. 783, 789--90 (1984).} In \textit{Calder}, the Supreme Court upheld California jurisdiction over Florida defendants who wrote and edited a defamatory article about a California resident, because they knew the article would harm the plaintiff in California and intended to exploit the California market.\footnote{\textit{Id.}}

Applied to grassroots digital speech acts: If Alice in Country A creates defamatory content about Bob in Country B, Country B courts may have jurisdiction over Alice under \textit{Calder} if she knew the content would harm Bob in Country B and expressly aimed the statement at Country B (e.g., by addressing it to Bob's Country B community). But the Supreme Court has limited \textit{Calder}: the defendant must engage in wrongful conduct ``expressly aimed'' at the forum, not merely cause foreseeable effects there.\footnote{\textit{Walden v. Fiore}, 571 U.S. 277, 287--90 (2014) (rejecting personal jurisdiction based solely on plaintiff's forum contacts).}

This creates uncertainty for grassroots forwarders. If Carol in Country C forwards Alice's defamatory content to David in Country B, does Country B have jurisdiction over Carol? Under \textit{Walden}, probably not, unless Carol expressly aimed the forwarding at Country B (knew David was in Country B, intended to harm Bob there, etc.). But if Carol forwards to a community spanning multiple countries, she may be subject to jurisdiction in any country where the content causes harm—a potentially global exposure.

\textbf{The amplification problem.} Viral spread through grassroots forwarding creates cascading jurisdictional exposure. If Alice's content is forwarded through a chain spanning dozens of countries, each forwarder might face jurisdiction in the country where the original plaintiff resides, plus any countries where they forwarded the content. Jurisdictional exposure therefore expands alongside dissemination itself: each additional forwarding event potentially creates new legal connections between participants, content, and multiple legal systems.

\textbf{No simplified jurisdictional framework.} The elimination of platform intermediaries makes personal jurisdiction more complex, not less. Each creator and forwarder must analyze whether their speech creates minimum contacts with foreign jurisdictions, whether the effects test applies, and whether their conduct was expressly aimed at the forum. This analysis must occur for every forwarding decision, not once by a central platform.

\subsubsection{Choice of Law in Cross-Border Speech}

Even when a court has jurisdiction, which country's law applies to the speech?

\textbf{Defamation law varies dramatically by jurisdiction.} U.S. defamation law provides robust protections for defendants: public figures must prove actual malice, truth is an absolute defense, and the First Amendment constrains liability.\footnote{See \textit{New York Times Co. v. Sullivan}, 376 U.S. 254 (1964); \textit{Philadelphia Newspapers, Inc. v. Hepps}, 475 U.S. 767 (1986).} English defamation law historically imposed stricter liability, placing the burden on defendants to prove truth and permitting larger damages.\footnote{But see Defamation Act 2013, c. 26 (U.K.) (reforming English defamation law to require claimants to prove serious harm).} Many countries criminalize defamation, creating exposure unavailable in U.S. civil law.

When Alice in the U.S. creates content about Bob in the U.K., which law applies? If Carol in Germany forwards it to David in France, which law governs Carol's liability? Traditional choice-of-law analysis examines: (1) where the plaintiff's reputation was harmed (place of injury), (2) where the defendant acted (place of conduct), (3) which jurisdiction has the most significant relationship to the dispute, and (4) which jurisdiction's law the parties expected to apply.\footnote{See Restatement (Second) of Conflict of Laws \S\S~145, 150 (1971).}

For digital speech forwarded globally, these factors point in multiple directions. The plaintiff's reputation may be harmed in every country where the content appears. The defendant acted wherever they were physically located when forwarding. Multiple jurisdictions have relationships to the dispute. Parties rarely have clear expectations about which law applies to peer-to-peer forwarding.

\textbf{Forum shopping and the race to the bottom.} Plaintiffs can sue in jurisdictions with the most favorable defamation laws. A public figure unable to prevail under U.S. actual malice standards might sue in a jurisdiction with stricter liability. The grassroots architecture provides no mechanism to prevent this forum shopping—unlike platforms, which can sometimes invoke choice-of-law clauses in Terms of Service or argue for application of a single law globally.

\subsubsection{Copyright and the Berne Convention}

Copyright presents somewhat more certainty due to international harmonization through the Berne Convention.\footnote{Berne Convention for the Protection of Literary and Artistic Works, Sept. 9, 1886, as revised at Paris on July 24, 1971, and amended in 1979, S. Treaty Doc. No. 99-27 (1986).} The Convention establishes minimum standards for copyright protection in over 180 member countries, including automatic protection upon fixation and national treatment (foreign works receive the same protection as domestic works).\footnote{\textit{Id.} arts. 5(1)-(2).}

Digital speech acts created by persons in Berne Convention countries receive copyright protection in all member countries. If Alice in France creates a digital speech act, it is automatically protected in the U.S., Japan, Australia, and all other Berne countries. This provides a degree of international copyright harmonization absent in defamation law.

However, enforcement remains national and territorial. Copyright is a bundle of territorial rights, enforceable in each country separately.\footnote{See \textit{Subafilms, Ltd. v. MGM-Pathe Commc'ns Co.}, 24 F.3d 1088, 1097 (9th Cir. 1994) (en banc) (copyright infringement is territorial).} If Bob in the U.S. forwards Alice's copyrighted digital speech act, Alice must sue Bob under U.S. copyright law in U.S. courts. If Carol in Germany also forwards it, Alice must sue Carol under German copyright law in German courts. There is no single global copyright enforcement action—each territorial infringement requires separate litigation.

\textbf{Practical enforcement challenges.} A creator whose digital speech act is forwarded virally across dozens of countries faces impossible enforcement burdens. They must: identify infringing forwarders in each country, retain counsel in each jurisdiction, navigate each country's copyright law and procedural rules, and obtain separate judgments in each forum. The grassroots architecture provides provenance chains identifying forwarders, but converting that technical information into legal enforcement across borders remains prohibitively expensive for most individual creators.

\subsubsection{Regulation Without Platform Intermediaries}

A defining characteristic of contemporary internet regulation is its reliance on intermediaries. Legislatures frequently regulate online communications indirectly by imposing obligations on entities that host, distribute, recommend, monetize, or otherwise facilitate user-generated content. This approach reflects a practical reality: platform operators aggregate communications that would otherwise be dispersed among millions of individual speakers and, therefore, function as regulatory leverage points. Lawmakers often regulate the intermediary through which those speakers communicate, rather than regulating each speaker individually.

Many contemporary regulatory regimes reflect this intermediary-focused model. The European Union's Digital Services Act requires platforms to implement moderation systems, respond to notice-and-action requests, assess systemic risks, and provide transparency reports.\footnote{Digital Services Act, arts. 14--16, 20, 23, 24, 34--35.} Germany's Network Enforcement Act (NetzDG) similarly imposes obligations on social-media platforms to remove unlawful material within specified timeframes or face substantial fines.\footnote{Gesetz zur Verbesserung der Rechtsdurchsetzung in sozialen Netzwerken [NetzDG] [Network Enforcement Act], Sept. 1, 2017, BGBl. I at 3352 (Ger.).} Comparable approaches appear throughout contemporary internet governance, where compliance obligations are directed primarily toward centralized intermediaries rather than individual participants.

Grassroots architectures substantially alter this regulatory model. Because content is stored on participant devices and disseminated through peer-to-peer forwarding, many of the functions that contemporary regulation assigns to centralized intermediaries are distributed across participants themselves. As a result, regulatory frameworks built around platform obligations—such as moderation requirements, notice-and-action procedures, transparency reporting, and centralized enforcement mechanisms—cannot be applied in the same manner. Where harmful or unlawful material spreads through a grassroots network, regulators cannot rely upon a single intermediary as a point of intervention. Responsibility instead becomes dispersed among the individuals who create, forward, and receive content, raising difficult questions about attribution, enforcement, and accountability when communications propagate across large numbers of participants and multiple jurisdictions.

This does not mean grassroots architectures exist outside legal regulation. Rather, the locus of regulation shifts from centralized intermediaries to individual participants. Instead of imposing obligations on a platform operator, regulators must rely upon existing legal doctrines directed at creators, forwarders, and other actors within the network: civil liability, criminal sanctions, injunctions, and related remedies.

\subsubsection{Jurisdictional Complexity, Not Simplicity}

The grassroots architecture does not eliminate international jurisdictional problems; it redistributes them. Platform architectures concentrate enforcement around centralized intermediaries, whereas grassroots architectures distribute legal exposure across the participants who create and disseminate content. This distribution carries both advantages and disadvantages. It reduces reliance on centralized gatekeepers and limits opportunities for platform-level censorship, but it also complicates questions of personal jurisdiction, increases opportunities for forum shopping, and makes enforcement more difficult when communications propagate across large networks.

These tradeoffs reflect a broader tension between centralized and distributed models of governance. Whether one approach better serves free expression, accountability, and democratic values depends upon contested judgments regarding the relative importance of enforcement capacity, individual responsibility, and resistance to concentrated control.

\section{Addressing Legal Challenges}\label{sec:challenges}

The objections that digital speech acts do not qualify for copyright protection fall into several categories: algorithmic generation challenges, idea/expression and functionality concerns, merger doctrine, mechanical generation precedents, statutory limitations, and broader scholarly critiques of copyright expansion. We address each in turn.

\subsection{Human Authorship and Algorithmic Execution}\label{subsec:algorithmic}

The strongest contemporary challenge to copyrightability arises from copyright law's requirement of human authorship. Recent cases addressing artificial intelligence and non-human creators reaffirm that copyright protects only works attributable to human creative activity. In Thaler v. Perlmutter, the D.C. Circuit held that an artwork generated autonomously by an artificial intelligence system could not receive copyright protection because the work lacked a human author. The Supreme Court subsequently denied certiorari, leaving that interpretation of the Copyright Act undisturbed.\footnote{\label{fn:thaler}
\textit{Thaler v. Perlmutter},
687 F. Supp. 3d 140 (D.D.C. 2023),
\textit{aff'd},
130 F.4th 1039 (D.C. Cir. 2025),
cert. denied,
No. 25-449 (U.S. Mar. 2, 2026).} The court emphasized that ``United States copyright law protects only works of human creation.''\footnote{\textit{Id.} at 146.} Likewise, in \textit{Naruto v. Slater}, the Ninth Circuit concluded that photographs taken by a macaque monkey could not be copyrighted because copyright law presupposes human authorship.\footnote{\label{fn:naruto}
\textit{Naruto v. Slater},
888 F.3d 418 (9th Cir. 2018).} The U.S. Copyright Office has adopted a similar position in its recent guidance on artificial intelligence, explaining that copyright protection depends upon the presence of human creative control over the expressive elements of a work.\footnote{\label{fn:ai-guidance}
U.S. Copyright Office, \textit{Copyright and Artificial Intelligence: Part 2 -- Copyrightability} (Jan. 2025), https://www.copyright.gov/ai/ai\_part\_2.pdf.}

At first glance, digital speech acts appear to raise a similar concern. A person creates expressive content, a cryptographic system records authorship and provenance information, and a signing algorithm generates a signature that binds those elements together. Because part of this process involves automated computational operations, one might argue that the resulting work lacks the human authorship required by copyright law.

The analogy, however, is ultimately misplaced because it focuses on the mechanism that authenticates expression rather than the source of the expression itself. The critical question under copyright law is not whether a machine participates in producing a work, but whether the expressive elements embodied in the work originate in human decision-making. This was the central insight of \textit{Burrow-Giles}. Faced with a photographic process that mechanically and chemically captured images, the Supreme Court did not ask whether the camera created the photograph. Instead, it examined whether the expressive features of the photograph originated with the photographer. Because Sarony selected the subject, pose, costume, lighting, and composition, the Court concluded that the resulting image embodied human authorship notwithstanding the camera's mechanical operation.\footnote{\textit{Burrow-Giles},
\textit{supra} note~\ref{fn:burrow-giles},
at 60.}

Digital speech acts present the same analytical structure. The person creating the digital speech act determines what is communicated, when it is communicated, under which identity it is communicated, and the communicative significance of the act itself. The cryptographic mechanisms do not generate those expressive choices; they authenticate and preserve them. Just as a camera fixed Sarony's creative decisions in photographic form, cryptographic processes fix and verify the author's expression within an attributable and tamper-evident record.

The analysis becomes clearer when the relevant work is properly identified. The copyrightable work is not the bare signature string viewed in isolation. It is the digital speech act as a whole: the expressive content, the cryptographic signature that authenticates it, and the metadata that situates it in time and identifies its speaker. Because the expressive elements of that work are selected and controlled by the human participant, the resulting work embodies human authorship notwithstanding the use of cryptographic processes.

This conclusion is consistent with the Copyright Office's distinction between AI-generated works and AI-assisted works. Copyright protection is unavailable where computational systems determine the expressive content of the work, but remains available where human beings exercise creative control and computational processes merely facilitate execution or implementation.\footnote{Copyright Office AI Guidance, \textit{supra} note~\ref{fn:ai-guidance}, at 9--13.} Digital speech acts fall squarely within the latter category because the algorithm authenticates and attributes expression rather than creating it.

\subsection{Idea, Expression, and Method Under \textit{Baker v. Selden}}\label{subsec:expression}

Section 102(b) of the Copyright Act codifies a foundational principle of copyright law: copyright protects expression, not ``any idea, procedure, process, system, method of operation, concept, principle, or discovery,'' regardless of the form in which such matters are embodied.\footnote{17 U.S.C. \S~102(b).} The Supreme Court articulated this distinction in \textit{Baker v. Selden}, holding that copyright in a book describing a bookkeeping system did not extend to the accounting method itself or to the forms necessary for its operation.\footnote{\label{fn:baker-selden}
\textit{Baker v. Selden},
101 U.S. 99 (1879).}

Selden's book explained a novel accounting method and included forms illustrating how the method was to be used. When Baker copied the forms, Selden's estate argued that copyright in the book extended to the accounting system embodied within it. The Supreme Court rejected that argument. Copyright protected Selden's explanation of the method, but not the method itself. As the Court explained, ``[t]he description of the art in a book, though entitled to the benefit of copyright, lays no foundation for an exclusive claim to the art itself. The object of the one is explanation; the object of the other is use.''\footnote{\textit{Id.} at 105.}

The significance of \textit{Baker} is not merely that methods are excluded from copyright protection. Rather, the case establishes the analytical framework for distinguishing protected expression from the systems, procedures, and functional mechanisms through which expression may be created or communicated. Applied to digital speech acts, the relevant question is therefore not whether cryptographic signature systems are copyrightable. They are not. The question is whether a digital speech act constitutes protected expression created through the application of such a system, or whether it is merely an instance of the system's operation.

The distinction becomes clearer when the work is properly characterized. Copyright does not protect the cryptographic signature algorithm itself. The algorithm is a mathematical procedure for authenticating content and therefore falls comfortably within §102(b)'s exclusion of procedures, processes, and methods of operation.\footnote{See \textit{Lotus}, \textit{supra} note~\ref{fn:lotus}, at 815 (holding menu command hierarchy was uncopyrightable method of operation).} Nor does copyright protect the general concept of binding content to identity through cryptographic means. These functional and conceptual elements remain free for all to use.

The digital speech act, however, is not the method. It is the expressive work produced through application of the method to particular human-authored content. In this respect, the relationship between a digital speech act and a cryptographic signature algorithm resembles the relationship between a photograph and a camera. Copyright does not protect the camera as a technology; it protects the photograph produced through the photographer's creative choices. Likewise, copyright does not protect the signature algorithm. It protects the resulting work when a person applies that method to content they have chosen, at a time they have chosen, and under an identity they have chosen to employ.

Modern software cases reinforce this distinction. In \textit{Lotus v. Borland}, the First Circuit held that a menu command hierarchy constituted an uncopyrightable ``method of operation'' because protecting it would effectively grant exclusive control over the way users operated the software.\footnote{\textit{Id.} at 816.} Similarly, in \textit{Computer Associates v. Altai}, the Second Circuit developed the abstraction--filtration--comparison framework to distinguish protectable expression from unprotectable ideas, processes, methods, and functional requirements embedded within software.\footnote{\label{fn:computer-assocs}
\textit{Computer Assocs. Int'l, Inc. v. Altai, Inc.},
982 F.2d 693, 706--11 (2d Cir. 1992).}

Applying the logic of \textit{Altai} to digital speech acts produces a straightforward result. At a high level of abstraction, a digital speech act implements the idea of authenticating content through a cryptographic signature. That idea is not copyrightable. The filtration step removes the signature algorithm itself as an unprotectable mathematical method, the concept of identity binding as an unprotectable idea, and the authentication function as an unprotectable method of operation. What remains is the person's expressive content together with the deliberate choice to authenticate that content under a particular identity at a particular moment. Although the authentication function may require that a signature exist, it does not dictate what content will be authenticated, when authentication will occur, or which identity will be associated with it. Those choices originate with the person creating the digital speech act.\footnote{\textit{Id.} at 709--10.}

The same analysis explains why the merger concerns underlying \textit{Baker} are absent here. The bookkeeping forms in \textit{Baker} merged with the accounting method because they were necessary to its use.\footnote{\textit{Baker},
\textit{supra} note~\ref{fn:baker-selden},
at 103.} By contrast, copyright in a particular digital speech act does not prevent others from authenticating content, claiming authorship, or employing cryptographic signatures. Multiple persons remain free to create their own digital speech acts using the same methods and technologies. Protecting one person's signed expression therefore does not confer exclusive rights over the underlying ideas or methods.

Properly understood, \textit{Baker v. Selden} supports rather than undermines copyrightability. The case directs courts to distinguish expressive works from the systems through which they are produced. The cryptographic signature method remains outside copyright's domain. The digital speech act created through application of that method remains within it.

\subsection{Merger Doctrine and Input-Dictated Expression}\label{subsec:merger}

The merger doctrine prevents copyright protection when idea and expression become inseparable. Where there are so few ways to express an idea that protecting a particular expression would effectively grant exclusive rights over the idea itself, copyright protection must yield to §102(b)'s exclusion of ideas from the copyright monopoly.\footnote{\label{fn:morrissey}
See \textit{Morrissey v. Procter \& Gamble Co.}, 379 F.2d 675, 678--79 (1st Cir. 1967) (``[W]hen there is essentially only one way to express an idea, the idea and its expression are inseparable and copyright is no bar to copying that expression.'').}

The merger question arises in a distinctive form for digital speech acts. Unlike many expressive works, a cryptographic signature is computationally determined by its inputs. Given particular content, a particular private key, a particular cryptographic algorithm, and a particular moment of signing, only one valid signature can result. This raises the concern that the signature may be dictated by functional constraints in much the same way that certain software declarations or technical specifications are dictated by system requirements.\footnote{See \textit{Oracle Am., Inc. v. Google Inc.}, 750 F.3d 1339, 1362--63 (Fed. Cir. 2014) (method declarations dictated by Java language requirements merged with idea), rev\textquotesingle d on other grounds, 141 S. Ct. 1183 (2021).}

Properly framed, however, the merger inquiry does not ask whether a signature is mathematically determined once particular inputs are selected. The relevant question is whether protecting the resulting work would effectively monopolize the underlying idea. The doctrine is concerned with preserving public access to ideas, methods, and functional systems, not with denying protection whenever a technological process produces a determinate result.

As discussed above, the copyrightable work is not the bare signature string viewed in isolation. It is the digital speech act as a whole: expressive content bound to identity and time through cryptographic authentication. The person's choices regarding what to communicate, when to communicate it, and under which identity to claim it constitute the expressive elements of the work. The signature serves to authenticate those choices rather than replace them.

The software cases reinforce this distinction. In \textit{Oracle}, the method declarations at issue were dictated by the requirements of the Java programming language itself.\footnote{\textit{Id.} at 1362--63.} Programmers seeking to invoke particular Java functions had no practical alternative but to employ the specified declarations. Protecting those declarations therefore threatened to restrict access to the underlying functionality. By contrast, the expressive elements of a digital speech act are not dictated by the authentication system. A person remains free to choose different content, different wording, different timing, and different identities. Although the resulting signature follows deterministically from those choices, the choices themselves remain matters of human expression rather than functional necessity.

The same analysis explains why the merger concerns underlying \textit{Baker} are absent here. The bookkeeping forms merged with the accounting system because they were necessary to its operation.\footnote{\textit{Baker},
\textit{supra} note~\ref{fn:baker-selden},
at 103.} Protecting the forms would therefore have granted exclusive rights over the method itself. By contrast, copyright in a particular digital speech act does not prevent others from authenticating content, claiming authorship, associating expression with identity, or employing cryptographic signatures. Multiple persons remain free to create their own digital speech acts using the same methods and technologies. Protection of one person's authenticated expression therefore does not confer exclusive rights over the underlying ideas or methods.

Merger doctrine therefore does not bar copyright in digital speech acts. The doctrine prohibits protection only when expression is so constrained that copyright would effectively confer ownership of an idea, method, or system. Digital speech acts do not create such a monopoly. The expressive choices embodied in the work remain distinct from the underlying concepts of authentication, attribution, and identity, and protection of those choices leaves the underlying ideas free for all to employ.

\subsection{Mechanical Reproduction and the Requirement of Originality}\label{subsec:mechanical}

A separate line of authority denies copyright protection to works produced through mechanical processes when those processes merely reproduce preexisting expression without the addition of human creative judgment. These cases are not principally concerned with technology. They address a more fundamental question: whether the purported author contributed any original expression beyond faithful reproduction of an existing work.

The leading cases involve technologically sophisticated reproductions whose objective was fidelity rather than authorship. In \textit{Bridgeman Art Library v. Corel Corp.}, exact photographic reproductions of public-domain paintings lacked the originality necessary for copyright protection because they added no new expressive elements.\footnote{\label{fn:bridgeman}
\textit{Bridgeman Art Library, Ltd. v. Corel Corp.},
36 F. Supp. 2d 191 (S.D.N.Y. 1999).} The same principle appeared in \textit{Meshwerks, Inc. v. Toyota Motor Sales U.S.A., Inc.}, where highly detailed digital wire-frame models of Toyota vehicles were denied protection because they reflected exact replication rather than creative interpretation.\footnote{\label{fn:meshwerks}
\textit{Meshwerks, Inc. v. Toyota Motor Sales U.S.A., Inc.},
528 F.3d 1258 (10th Cir. 2008).} Likewise, in \textit{ATC Distribution Group v. Whatever It Takes Transmissions \& Parts, Inc.}, the Sixth Circuit treated catalog photographs designed to depict products accurately as lacking the originality required for copyright protection.\footnote{\label{fn:atc}
\textit{ATC Distrib. Grp., Inc. v. Whatever It Takes Transmissions \& Parts, Inc.},
402 F.3d 700 (6th Cir. 2005).} Taken together, these decisions establish a narrow but important proposition: copyright does not attach merely because a technologically sophisticated process produces an output. When a process serves only to reproduce an existing work with fidelity and leaves no room for meaningful creative judgment, the resulting output lacks the originality required by copyright law.\footnote{
\textit{Bridgeman},
\textit{supra} note~\ref{fn:bridgeman};
\textit{Meshwerks},
\textit{supra} note~\ref{fn:meshwerks},
at 1265--66;
\textit{ATC},
\textit{supra} note~\ref{fn:atc},
at 707.
}

Digital speech acts do not fall within this category. Unlike the works at issue in \textit{Bridgeman}, \textit{Meshwerks}, and \textit{ATC}, they are not reproductions of preexisting expression. The person creating the digital speech act determines the expressive content, the moment of publication, the identity under which the expression is asserted, and the communicative significance of the act itself. Those choices precede the operation of the cryptographic system and supply the originality embodied in the work. The cryptographic signature does not reproduce expression; it authenticates expression already chosen by the human participant. The algorithm therefore functions as a mechanism of attribution and verification rather than as a substitute for creative judgment.

Moreover, it is not necessary to argue that the signature string, considered in isolation, possesses independent originality. As discussed above, the relevant work is the digital speech act as a whole: expressive content bound to a cryptographic signature and associated metadata. Originality resides in the human-authored expression and the human decisions embodied in the act. The signature contributes authentication and attribution, not independent creative content.

Accordingly, the mechanical-reproduction cases do not undermine copyrightability. They deny protection where technology merely replicates preexisting expression without creative intervention. Digital speech acts involve the opposite relationship. Human authors make the expressive choices, while cryptographic systems serve only to preserve and verify those choices. The originality required by copyright law therefore remains attributable to the person rather than to the technology.

\subsection{Thin Copyright, Overclaiming, and the Limits of Protection}

Even where a work satisfies the requirements of authorship, originality, and fixation, copyright protection may remain problematic if the resulting rights facilitate strategic overclaiming. The literature on ``thin copyright'' raises precisely this concern: works resting on only minimal originality may technically qualify for protection while nevertheless creating anticommons effects, impeding follow-on creation, or enabling attempts to control functional and informational subject matter beyond copyright's proper domain.

Mark Lemley has argued that minimally creative works can produce forms of ``thin copyright'' that encourage strategic overclaiming.\footnote{Lemley, \textit{supra} note~\ref{fn:lemley-property}, at 1059--65.} Copyright holders may attempt to leverage protection in genuinely creative elements to claim effective control over functional, factual, or otherwise unprotectable material, thereby generating anticommons problems that impede later creators.\footnote{\textit{Id.}} Pamela Samuelson has similarly cautioned against applying copyright's minimal originality threshold so expansively that formulaic or functionally determined expressions receive protection beyond copyright's proper domain.\footnote{Samuelson, \textit{supra} note~\ref{fn:samuelson-originality}.}

These concerns merit consideration because digital speech acts combine expression with authentication infrastructure. The question is whether recognizing copyright in such works would permit authors to exercise control over authentication systems, network functions, or other functional elements of digital communication.

Both the architecture of grassroots systems and the structure of copyright doctrine substantially limit that possibility. As Part~\ref{sec:core} demonstrated, the copyright claim advanced in this Article is confined to expressive digital speech acts rather than the full range of cryptographically authenticated transactions. Functional digital speech acts—such as votes, token transfers, and similar operational acts—derive their force from cryptography, contract, and the digital social contract rather than copyright. The thin-copyright concerns identified by Lemley and Samuelson therefore arise only with respect to expressive works, where copyright remains confined to particular acts of expression rather than the functional architecture of digital communication.\footnote{\textit{Feist}, \textit{supra} note~\ref{fn:feist}, at 345.}

This limitation also preserves the distinction between expressive works and the authentication infrastructure through which they are created. Recognition of copyright in expressive digital speech acts does not grant exclusive rights over cryptographic algorithms, authentication procedures, verification processes, or key-management systems. Those systems remain analytically distinct from the resulting work and continue to operate independently of copyright protection.\footnote{Lemley, \textit{supra} note~\ref{fn:lemley-property}, at 1062--65.}

The limited scope of the copyright claim is reinforced by established copyright doctrine. As Part~\ref{sec:challenges}.\ref{subsec:merger} demonstrated, merger doctrine denies protection where idea and expression become inseparable.\footnote{See \textit{Morrissey}, \textit{supra} note~\ref{fn:morrissey}, at 678--79.} Elements dictated by functional requirements or computational necessity therefore remain outside copyright's reach. Section 102(b) provides a second and independent limitation by excluding from protection ``any idea, procedure, process, system, method of operation, concept, principle, or discovery'' regardless of the originality embodied in surrounding expression.\footnote{17 U.S.C. §~102(b).} The authentication functions of cryptographic systems, the procedures through which signatures are generated and verified, and the broader architecture of identity binding therefore remain categorically outside the scope of copyright.

Accordingly, these limitations prevent the forms of strategic overclaiming identified by Lemley and Samuelson. Merger doctrine and §102(b) exclude functional elements, methods, systems, and processes from protection. What remains protectable is the person's particular expressive act of communication, fixed, attributed, and authenticated through cryptographic means. Recognition of copyright in such works therefore does not create the anticommons concerns associated with thin copyright, because neither the architecture of grassroots systems nor the structure of copyright doctrine permits exclusive control over the underlying methods of authentication, communication, or network participation.

\subsection{Other Statutory and Policy Objections}

\subsubsection{Contract Preemption Under Section 301}

Section 301 preempts state-law rights that are equivalent to copyright's exclusive rights.\footnote{17 U.S.C. \S~301(a).} Courts generally apply the ``extra element'' test, under which a state-law claim survives preemption when it requires proof of an element beyond those necessary to establish copyright infringement.\footnote{\textit{Computer Assocs.},
\textit{supra} note~\ref{fn:computer-assocs},
at 716.} Because contractual obligations arise from voluntary agreement rather than from the mere act of copying, distributing, or displaying a work, courts have consistently treated contract claims as containing the requisite extra element.\footnote{\textit{ProCD},
\textit{supra} note~\ref{fn:procd},
at 1454--55 (contract claim not preempted because it requires proof of extra element---agreement---beyond copyright infringement).}

Digital social contracts fall comfortably within this framework. The non-unbundling obligations described in Part~\ref{sec:grassroots} do not arise from copyright ownership itself. They arise from agreements among participants concerning the treatment of digital speech acts within a particular network. A participant who strips signatures, destroys provenance information, or falsifies attribution violates those obligations because the participant agreed not to do so, not merely because the participant copied or distributed a copyrighted work. The requirement of assent therefore supplies the extra element that takes such claims outside §301 preemption.\footnote{\textit{Computer Assocs.},
\textit{supra} note~\ref{fn:computer-assocs},
at 716;
\textit{ProCD},
\textit{supra} note~\ref{fn:procd},
at 1454--55.}

The interests protected by digital social contracts are also distinct from those protected by copyright. Copyright secures exclusive rights in reproduction, distribution, adaptation, performance, and display. Digital social contracts instead govern attribution, provenance preservation, and cryptographic integrity within particular communities. A participant may comply fully with copyright law while violating contractual obligations concerning signatures or provenance, just as a participant may infringe copyright without breaching any contractual commitment. The two regimes therefore regulate different conduct and protect different interests.

Attribution obligations warrant brief separate consideration because copyright law itself recognizes limited attribution and integrity interests through the Visual Artists Rights Act.\footnote{17 U.S.C. \S~106A.} VARA, however, applies only to a narrow category of works of visual art.\footnote{17 U.S.C. \S~101 (defining ``work of visual art'').} Digital speech acts generally fall outside that category. Contractual commitments to preserve signatures and provenance therefore do not duplicate the attribution rights Congress created through VARA. Moreover, attribution-related claims frequently involve additional elements such as deception, fraud, or false designation, which further distinguish them from ordinary copyright claims.\footnote{See \textit{Dastar Corp. v. Twentieth Century Fox Film Corp.}, 539 U.S. 23 (2003) (Lanham Act false designation claim involves elements beyond copyright).}

Digital social contracts and copyright therefore perform complementary rather than equivalent functions. Copyright establishes ownership of digital speech acts and provides remedies against unauthorized exploitation by third parties. Digital social contracts establish consensual obligations governing attribution, provenance, and non-unbundling among participants. Because those obligations depend upon agreement and protect interests distinct from the exclusive rights enumerated in §106, they are not preempted by §301.\footnote{\textit{Computer Assocs.},
\textit{supra} note~\ref{fn:computer-assocs},
at 716;
\textit{ProCD},
\textit{supra} note~\ref{fn:procd},
at 1454--55.}

\subsubsection{First Sale Doctrine and the Limits of Section 109}

Section 109(a) provides that ``the owner of a particular copy or phonorecord lawfully made under this title . . . is entitled, without the authority of the copyright owner, to sell or otherwise dispose of the possession of that copy or phonorecord.''\footnote{17 U.S.C. \S~109(a).} The first-sale doctrine therefore limits a copyright owner's ability to control downstream transfers of a lawfully acquired copy. Once ownership of a particular copy has passed, the copyright owner may not invoke the distribution right to prevent that copy from being resold, lent, or otherwise transferred.

The doctrine does not, however, confer a general right to alter a copyrighted work, create modified versions of it, or disregard independent contractual obligations governing its use. Most fundamentally, §109 concerns disposition of ``that copy'' which the recipient owns.\footnote{17 U.S.C. \S~109(a) (emphasis added).} It permits transfer of a lawfully acquired copy, but does not authorize conduct that implicates rights beyond distribution.

This distinction is particularly important because the copyright claim advanced throughout this Article treats the digital speech act as the relevant work. The work consists not merely of semantic content in isolation, but of content as authenticated, attributed, and temporally situated through the act of signing. A recipient may transfer the digital speech act as received, including its associated signatures and provenance information. Removing those elements, however, does not simply transfer an existing copy; it creates a different object than the one originally received. To the extent such conduct constitutes reproduction or preparation of a modified work, it implicates rights that first sale does not extinguish.\footnote{See 17 U.S.C. \S~109(a) (preserving rights beyond the limited transfer privilege recognized by the doctrine).}

The limits of first sale become even more pronounced in digital environments, where redistribution ordinarily occurs through copying rather than through physical transfer of an existing object. In \textit{Capitol Records, LLC v. ReDigi Inc.}, the Second Circuit held that a digital music resale service fell outside first-sale protection because transferring digital files necessarily involved reproduction rather than mere distribution of the original copy.\footnote{\textit{Capitol Records, LLC v. ReDigi Inc.}, 910 F.3d 649, 657--60 (2d Cir. 2018).} The doctrine protected disposition of a particular copy; it did not authorize creation of a new one.\footnote{\textit{Id.}} Digital speech acts present a similar issue. Forwarding a signed expression ordinarily creates additional copies on additional devices rather than transferring possession of a single physical artifact. To the extent redistribution requires reproduction, the conduct falls outside the traditional scope of §109 even before questions of attribution or provenance arise.\footnote{\textit{Id.}}

Nor does first sale displace the contractual dimension of grassroots architectures. The doctrine limits copyright claims; it does not invalidate contractual obligations voluntarily assumed by participants. Although \textit{Bobbs-Merrill Co. v. Straus} established first sale as a limit on copyright owners' ability to control downstream resale, it did not eliminate the possibility of contractual restrictions operating independently of copyright law.\footnote{\textit{Bobbs-Merrill Co. v. Straus}, 210 U.S. 339, 350--51 (1908).} Likewise, \textit{ProCD, Inc. v. Zeidenberg} recognized that contractual use restrictions may coexist with copyright limitations because contractual obligations arise from agreement rather than from exclusive rights enforceable against the world.\footnote{\textit{ProCD},
\textit{supra} note~\ref{fn:procd},
at 1454.}

Digital social contracts operate in precisely this manner. Their purpose is not to expand copyright's exclusive rights but to establish mutually accepted obligations concerning attribution, provenance preservation, and non-unbundling. A participant who agrees not to strip signatures may remain bound by that commitment even where copyright law alone would not prohibit a particular transfer. The source of the obligation is contractual rather than proprietary.

Accordingly, first sale does not undermine the theory of copyright advanced here. Section 109 permits redistribution of lawfully acquired copies, but it does not create a right to sever authorship information, eliminate provenance, reproduce works during transmission, or disregard contractual commitments concerning attribution. The doctrine limits copyright's distribution right while leaving intact the contractual obligations through which grassroots systems preserve the integrity of digital speech acts.

\subsubsection{Scholarly Critiques and the Scope of Copyright}

Copyright scholars have long cautioned against extending copyright beyond its traditional domain. Jessica Litman, James Boyle, and others have argued that many modern expansions of copyright have produced additional private rights without corresponding public benefits, often serving incumbent interests at the expense of access, competition, and follow-on creation.\footnote{
Jessica Litman, \textit{Digital Copyright} (2001);
\label{fn:boyle}
James Boyle, \textit{The Public Domain: Enclosing the Commons of the Mind} (2008).
} From this perspective, recognizing copyright in digital speech acts might appear to be another step in copyright's continual expansion into new technological environments.\footnote{Boyle, \textit{supra} note~\ref{fn:boyle}, at 54--97.}

That characterization, however, depends upon treating digital speech acts as a new category of copyrightable subject matter. The argument advanced here is narrower. Digital speech acts are not proposed as a new class of works. Rather, they are expressions that satisfy the existing requirements of authorship, originality, and fixation. Copyright protection therefore arises not from expanding copyright's boundaries but from applying established doctrine to expressions created within a new technical architecture.\footnote{17 U.S.C. \S~102(a).}

A related critique concerns incentives. Copyright is traditionally justified as a mechanism for encouraging creative production,\footnote{See U.S. Const. art. I, § 8, cl. 8 (granting Congress power ``[t]o promote the Progress of Science'' through exclusive rights).} yet individuals routinely create messages, photographs, comments, and social-media posts without regard to copyright incentives.\footnote{See Yochai Benkler, \textit{Sharing Nicely: On Shareable Goods and the Emergence of Sharing as a Modality of Economic Production}, 114 Yale L.J. 273 (2004).} But the argument advanced here does not depend primarily on incentivizing additional expression. Rather, copyright supports the ability of participants in grassroots architectures to retain ownership of the expressions they create instead of relinquishing effective control to centralized intermediaries.\footnote{See Neil Weinstock Netanel, \textit{Copyright's Paradox} 53--82 (2008) (describing copyright's functions beyond simple incentives to create).}

Other scholars have questioned whether automatic copyright protection grants legal rights to works whose creators neither need nor desire protection.\footnote{See Christopher Sprigman, \textit{Reform(aliz)ing Copyright}, 57 Stan. L. Rev. 485 (2004).} Digital speech acts partially address this concern through architecture. Participation in a cryptographically authenticated system requires deliberate acts of identification and signing, and grassroots systems may incorporate licensing preferences, attribution requirements, and provenance rules directly into their operation.\footnote{\textit{Id.} at 490--505.}

Platform-power critiques likewise merit attention. Copyright in user-generated content has often benefited intermediaries more than speakers because platforms obtain broad licenses or transfers through Terms of Service and exploit those rights at scale.\footnote{See Tarleton Gillespie, \textit{Wired Shut: Copyright and the Shape of Digital Culture} 137--69 (2007).} If digital speech acts merely generated additional rights for platforms to appropriate, the critique would be persuasive. The architecture examined here, however, distributes both ownership and possession among participants rather than centralized intermediaries. Copyright therefore strengthens the legal position of speakers rather than facilitating intermediary control.\footnote{See Boyle, \textit{supra} note~\ref{fn:boyle}, at 54--77 (documenting copyright expansion benefiting media companies).}

Finally, scholars have expressed concern that copyright protection for political and social expression may burden democratic discourse.\footnote{See Rebecca Tushnet, \textit{Copy This Essay: How Fair Use Doctrine Harms Free Speech and How Copying Serves It}, 114 Yale L.J. 535 (2004).} Existing doctrine already contains substantial safeguards. Fair use protects criticism, commentary, and transformative engagement with copyrighted expression,\footnote{See \textit{Campbell v. Acuff-Rose Music, Inc.}, 510 U.S. 569 (1994).} while the idea-expression distinction and fair-use doctrine together form part of the constitutional accommodation between copyright and free expression.\footnote{See \textit{Eldred v. Ashcroft}, 537 U.S. 186, 219--20 (2003).} Within grassroots architectures, copyright primarily supports attribution and resistance to unauthorized appropriation while leaving substantial space for circulation, criticism, and democratic engagement.

These critiques identify genuine concerns, but they do not establish that copyright is inappropriate for digital speech acts. Much of the skepticism surrounding copyright expansion assumes centralized ownership, intermediary control, and concentration of rights. The architecture examined here presents a different configuration in which ownership, possession, and authorship remain aligned with individual speakers. Whether that arrangement ultimately produces better social outcomes remains an empirical question. The narrower claim advanced in this Article is simply that existing copyright doctrine is capable of recognizing digital speech acts without producing the forms of concentration and appropriation that motivate many contemporary critiques of copyright expansion.

\subsection{From Copyrightability to Property and Self-Governance}

Taken together, these doctrines leave the traditional boundaries of copyright intact while permitting recognition of digital speech acts as copyrightable works. Digital speech acts remain expressions created through human volition, fixed in tangible media, and sufficiently original to satisfy copyright's threshold for protection. The cryptographic mechanisms through which they are authenticated do not eliminate authorship, transform expression into process, or convert creative acts into purely mechanical reproductions.\footnote{See Parts~\ref{sec:challenges}.\ref{subsec:algorithmic},\ref{subsec:expression}, \ref{subsec:mechanical}.} Existing doctrines such as merger and §102(b) continue to limit protection where functional considerations predominate\footnote{Parts~\ref{sec:challenges}.\ref{subsec:merger}}, while doctrines governing thin copyright, preemption, and first sale constrain the scope of any rights recognized. The result is not an expansion of copyright beyond its traditional boundaries, but an application of established principles to a new technological environment.

This conclusion does not resolve every question. Courts may disagree about the significance of cryptographic authentication, the level of creativity required for particular digital speech acts, or the broader policy implications of recognizing copyright in decentralized communications. The relevant inquiry, however, is not whether uncertainty exists, but whether existing doctrine forecloses recognition of copyright in digital speech acts. It does not.

Recognition of copyright in digital speech acts is itself a consequential doctrinal conclusion, with implications for contemporary understandings of copyright, digital authorship, and online participation. Yet its importance extends beyond copyright doctrine. Once authorship, attribution, ownership, and possession of digital expressions can coalesce within the same expressive act, copyright becomes relevant to broader questions of property, governance, and democratic participation. The legal recognition of digital speech acts therefore provides a foundation for examining how digital communities organize, coordinate, and distribute power.
 \section{Property and Democratic Self-Governance}\label{sec:democratic-foundations}

Political theorists across traditions have long recognized that property is not merely an economic institution but a democratic one. The central concern is not wealth accumulation as such, but independence. Citizens who possess productive assets participate in social, economic, and political life on different terms than citizens whose participation depends upon the continuing permission of others. Ownership provides a sphere of autonomy within which individuals can act, create, associate, and speak without being wholly subject to external control.

These concerns acquire particular significance in digital environments. Contemporary social, political, and economic participation increasingly occurs through digital systems, yet that participation is frequently mediated through centralized platform intermediaries that control the infrastructure upon which communication, association, and expression depend.  Grassroots architectures raise the possibility of a different arrangement. Whether such an arrangement advances democratic values depends upon how theories of property, ownership, and independence evaluate relationships of dependence and control within digital forms of participation.

\subsection{From Labor to Democratic Independence}

John Locke grounded property rights in self-ownership and labor.\footnote{Locke, \textit{supra} note~\ref{fn:locke}, §§ 25--51.} One owns oneself and therefore acquires claims in the products of one's labor.\footnote{\textit{Id.} \S~27.} Although Lockean labor theory remains influential, its application to intellectual property has been extensively criticized. Scholars have questioned whether labor alone can justify exclusion rights in non-rivalrous goods, whether creators can claim ownership over expressions built from common cultural resources, and whether intellectual property creates artificial scarcity rather than protecting genuine interests.\footnote{\label{fn:hettinger}Edwin C. Hettinger, \textit{Justifying Intellectual Property}, 18 Phil. \& Pub. Aff. 31, 35--38 (1989); \label{fn:gordon}Wendy J. Gordon, \textit{A Property Right in Self-Expression: Equality and Individualism in the Natural Law of Intellectual Property}, 102 Yale L.J. 1533, 1540--49 (1993); \label{fn:hughes}Justin Hughes, \textit{The Philosophy of Intellectual Property}, 77 Geo. L.J. 287, 296--330 (1988); \label{fn:shiffrin}Seana Valentine Shiffrin, \textit{Lockean Arguments for Private Intellectual Property}, in \textsc{New Essays in the Legal and Political Theory of Property} 138 (Stephen R. Munzer ed., 2001).} Those critiques weaken efforts to treat labor alone as a sufficient basis for ownership of digital speech acts. They do not, however, eliminate the question of who should own digital expressions or why that ownership matters. 

Democratic theories of property shift the focus from the origins of ownership claims to the consequences of ownership arrangements. Rather than asking whether digital participants deserve ownership because they labored, they ask how different ownership arrangements distribute power, structure relationships of dependence, and affect the conditions of democratic participation. 

This concern becomes particularly salient in digital environments. Social, political, and economic participation increasingly occurs through digital expression, through which individuals build reputations, form associations, organize collective action, create economic value, and engage in public discourse. Yet these activities typically occur through platform architectures in which intermediaries control the technical infrastructure of participation and obtain extensive contractual rights over user-created content.

The democratic significance of ownership over digital expressions therefore lies not in rewarding labor, but in structuring relationships of power. Ownership determines who controls participation, who captures value, and who possesses the legal authority to decide how digital expressions may be used. It is this question of distributed versus concentrated control that links theories of property to theories of democratic self-government.

Existing debates regarding digital democracy or platform governance frequently ask how individuals can retain greater control over information already collected by platforms. This Article asks a question \textit{a priori}: who controls the productive assets generated through digital participation in the first place? Questions of privacy, data governance, transparency, accountability, and content moderation largely arise after ownership and control have already been allocated. Democratic theory requires examination of the allocation itself.

\subsection{Digital Expression as Productive Assets}

Much contemporary scholarship concerning digital sovereignty frames ownership as a mechanism for protecting privacy, preserving autonomy, or increasing user control over digital systems. These concerns are important, but they remain largely reactive. They seek to constrain the power of existing intermediaries without addressing the underlying distribution of productive assets that gives rise to intermediary power in the first place. The argument advanced here is different. 

Digital expressions are not merely objects of privacy or autonomy interests. They increasingly function as productive assets through which individuals generate economic value, social capital, political influence, and participation in public life. 

The productive capacity of digital expression does not ordinarily arise from any single post, photograph, recording, or message considered in isolation. Rather, it emerges from the accumulation of attributed expressions over time. Through repeated acts of communication, individuals build creative portfolios, professional reputations, social relationships, audiences, and bodies of work whose value depends upon continuing attribution to their creator. It is this accumulated corpus of authenticated expression, together with the associated metadata, provenance, social relationships, and reputation that accrue through verified attribution, rather than any individual digital expression, that increasingly functions as a productive asset within digital society.

Among contemporary democratic theories of property, Rawls's account of property-owning democracy provides the strongest framework for understanding the democratic significance of ownership of digital speech acts.\footnote{Rawls, \textit{supra} note~\ref{fn:rawls-jf}, at 135--40.}

Rawls distinguished property-owning democracy from welfare-state capitalism on the basis of ownership. In welfare-state capitalism, productive assets remain concentrated in relatively few hands, while the majority participate primarily as employees or beneficiaries of redistribution. In a property-owning democracy, productive assets are widely dispersed throughout society, enabling citizens to participate as independent agents rather than dependents.\footnote{\textit{Id.} at 139.} As Rawls explained, the objective is to prevent a small segment of society from controlling economic life and thereby exercising indirect control over political life as well.\footnote{\textit{Id.}}

The central insight is structural. Democratic equality requires more than formal rights. It requires sufficiently dispersed ownership that citizens can participate in social and economic life without being subject to the arbitrary power of concentrated private institutions. Although Rawls wrote primarily with land, capital, and productive enterprises in mind, the underlying principle is not limited to physical assets. The relevant category is productive assets—resources that enable meaningful participation in the economy and society.\footnote{\label{fn:oneill-williamson}See Martin O'Neill \& Thad Williamson, \textsc{Property-Owning Democracy: Rawls and Beyond} (2012); Samuel Freeman, \textit{Capitalism in the Classical and High Liberal Traditions}, 28 Soc. Phil. \& Pol'y 19 (2011); \label{fn:williamson-realizing}Thad Williamson, \textit{Realizing Property-Owning Democracy: A 20-Year Strategy to Create an Egalitarian Distribution of Assets in the United States}, in O'Neill \& Williamson, \textit{supra}, at 225.}

The implication for digital environments follows directly. Because accumulated portfolios of attributed digital expression increasingly function as productive assets, ownership of those expressions affects whether individuals participate in digital life as independent actors or as dependents within systems controlled by others. Distributed ownership of digital expressions therefore serves the same objective that motivated Rawls's property-owning democracy: preserving the independence of citizens by dispersing ownership of the assets upon which participation depends.

The classification of digital expressions as productive assets is not theoretical speculation. Contemporary platform firms derive extraordinary economic value from user-generated content and associated social interactions. As discussed in Part~\ref{sec:platform-regime}, public filings from Meta and Alphabet reveal revenues measured in the hundreds of billions of dollars generated through systems built upon monetizing user expression, engagement, and participation. The economic significance of these activities demonstrates that digital expression has become an important productive resource within contemporary society. The question for democratic theory is therefore not whether such assets exist, but how ownership and control over them are distributed.

Rawls's concern was not with the particular form productive assets took, but with the democratic consequences of their concentration. In agrarian societies the relevant assets were land; in industrial societies they were factories and machinery. In digital societies, expressive and informational assets increasingly perform the same productive function. If ownership and control of those assets becomes concentrated within a small number of platform intermediaries, the structural conditions Rawls identified reappear in digital form.

This perspective reframes copyright ownership of digital speech acts. The question is not whether a single post, image, or message possesses substantial economic value. The question is whether the aggregate ownership and control of digital expression is distributed among persons or concentrated within intermediary institutions. Platform architectures concentrate control over digital participation. Users participate, but the productive assets generated through that participation—their accumulated portfolios of attributed digital expression—remain institutionally controlled.

\subsection{Ownership, Possession, and Distributed Power}

The democratic significance of ownership over digital expressions becomes clearer in grassroots architectures because ownership and possession coincide. Participants not only retain copyright in their digital expressions, but also possess the devices, data, and cryptographic credentials through which those expressions are created and transmitted. The combination of copyright ownership and cryptographic possession creates a form of digital independence that neither legal rights nor technical control could achieve alone. The copyrightability of digital speech acts therefore functions as one mechanism through which ownership of productive assets can remain broadly distributed in an increasingly digital society.

The significance of control over digital expressions extends beyond economics because ownership structures are ultimately governance structures. As Joseph Singer argues, property is not merely a relationship between persons and things, but a structure of relationships among persons.\footnote{Singer, \textit{supra} note~\ref{fn:singer-entitlement}.} Ownership therefore determines who possesses authority over whom and under what conditions participation occurs. 

Similar observations appear in scholarship on digital governance. As Luca Belli and Jamila Venturini observe, one of the primary functions of contracts is to fashion power relationships.\footnote{Luca Belli \& Jamila Venturini, \textit{Private Ordering and the Rise of Terms of Service as Cyber-Regulation}, 5 Internet Policy Rev. 1, 3 (2016).} Platform Terms of Service therefore function not merely as agreements but as instruments that allocate authority within digital environments. They determine who may speak, who may access audiences, who may monetize expression, and who may exclude others from participation.

Property and contract thus operate together to structure participation in digital life. Platform architectures create relationships of dependence because intermediaries control both the productive assets of participation and the contractual conditions under which participation occurs. Grassroots architectures can redistribute these relationships by returning ownership and possession to participants themselves, reducing dependence upon intermediary institutions and dispersing authority throughout the network. The result is not the elimination of governance, but the decentralization of governance authority itself.

\subsection{Democratic Ownership and Anti-Enclosure}

Commons-oriented scholars may object that democratic participation is better served through shared informational commons than through proprietary rights. That concern is important, particularly given the history of copyright expansion and enclosure of information resources. Yet the alternative examined in this Article is not a choice between commons and ownership in the abstract. It is a choice between ownership and control concentrated in platform intermediaries and ownership distributed among participants. Democratic alternatives—including commons-based governance, open licensing, cooperative production, and other forms of collective stewardship—depend upon participants retaining meaningful authority over the expressive assets they create. Grassroots architectures can preserve those possibilities by ensuring that copyright ownership remains with creators rather than intermediary institutions. Participants may choose to share, license, pool, or collectively govern their expressive assets, but those decisions remain theirs to make. The democratic concern is therefore not enclosure of the commons by individuals, but enclosure of the digital commons by platforms.

In this respect, copyright with regards to digital speech acts serves an anti-enclosure function. The right is not invoked to centralize control over expression, but to prevent the transfer of expressive ownership from creators to intermediary institutions.

\subsection{Distributed Ownership and Democratic Participation}

No single theory fully justifies ownership of digital speech acts. Lockean labor theory faces familiar objections. Rawls did not write about copyright. Singer's relational theory imposes obligations as well as rights. Commons scholars correctly warn against enclosure. Yet these traditions converge on a common concern: preventing domination through concentrated control of the resources necessary for participation. In digital society, expressive assets increasingly constitute such resources. Copyright ownership of digital speech acts thus serves democratic values not because it maximizes exclusion, but because it prevents the consolidation of expressive ownership within intermediary institutions. It disperses control, preserves attribution, and provides a legal foundation for digital participation independent of platform permission. 
\section{Economic Considerations }\label{sec:economic-implications}

Ownership of digital expression does not merely allocate control over participation; it allocates the capacity to capture economic value generated through participation. As discussed in Part~\ref{sec:platform-regime}, contemporary platform firms derive extraordinary revenues from systems built upon monetizing user expression, interaction, and engagement. Economic value generated through collective participation is therefore captured primarily at the platform layer.

Grassroots architectures alter this allocation. When participants retain ownership, possession, and control over their digital expressions, they also retain the ability to determine how the economic value associated with those expressions is realized. The significance of distributed ownership is therefore not limited to questions of privacy, autonomy, or democratic participation. It also concerns the distribution of economic agency within digital environments.

\subsection{Grassroots Architectures as Property-Owning Networks}

The preceding discussion demonstrated that digital expression increasingly function as productive assets within contemporary society. The economic implication of that conclusion is straightforward: ownership of digital speech acts determines who may capture the value those assets generate.

Rawls's account of property-owning democracy was concerned with dispersing productive assets throughout society rather than concentrating them within a small number of institutions. Grassroots architectures may be understood as extending that principle into the digital economy. Rather than concentrating ownership of expressive and informational assets within platform firms, they distribute ownership among the participants who create, possess, and maintain those assets.

This redistribution alters the economic structure of participation. Under platform architectures, users contribute content, attention, behavioral information, and social interaction to systems owned by others. Participation generates value, but ownership of the infrastructure and associated assets remains concentrated. Under grassroots architectures, people remain participants while simultaneously occupying the position of owners. They possess the expressive assets through which value is generated and retain the legal rights necessary to determine how those assets may be used.

The economic significance of this distinction extends beyond any particular revenue model. Copyright ownership of digital speech acts provides the legal foundation upon which future forms of exchange, licensing, cooperation, and commercialization may be built. Participants become economic actors rather than merely sources of platform-generated value.

\subsubsection{Economic Autonomy and Creator Commercialization}

The combination of possession and copyright ownership creates forms of economic autonomy unavailable within intermediary-controlled systems.

First, creators may license digital speech acts directly. Copyright ownership permits authors to authorize reproduction, distribution, adaptation, and other uses of their expressions. Grassroots architectures place those rights in the hands of participants rather than intermediary institutions. Creators may therefore negotiate directly with readers, viewers, publishers, applications, or other entities seeking access to expressive works.

Second, creators may engage in direct sales or subscription arrangements. Rather than relying upon platform-mediated subscription or advertising systems, participants may condition access to content on payment, membership, contractual participation, or other arrangements determined by the creator. Whether implemented through traditional payment systems, digital currencies, or cooperative structures, the legal authority to establish such arrangements derives from copyright ownership itself.

Third, creators may organize collectively. Data cooperatives and related governance structures have been proposed as mechanisms through which individuals can exercise collective control over informational resources.\footnote{See Salomé Viljoen, \textit{A Relational Theory of Data Governance}, 131 Yale L.J. 573 (2021).}
\footnote{See also Thomas Hardjono et al., \textit{Toward a Design Philosophy for Interoperable Blockchain Systems}, MIT Connection Science (2018).} Copyright ownership of digital speech acts creates similar possibilities for collective licensing, rights management, syndication, and cooperative bargaining. Rather than assigning rights to platforms, creators may pool rights while retaining collective control over their use.

Fourth, ownership preserves attribution and reputation. Economic value frequently arises not merely through direct payment but through recognition, audience development, professional opportunities, and social influence. Where attribution remains cryptographically bound to expression, participants retain the reputational value generated through their contributions rather than allowing that value to be absorbed into platform-controlled ecosystems.

These possibilities do not depend upon any single business model. The central point is more modest but more fundamental: ownership creates the legal capacity for creators to determine how value generated by their expressions will be captured and distributed.

\subsubsection{Intermediary Rent Extraction and Alternative Market Structures}

The concentration of ownership within platform architectures has broader consequences for market organization.

Platform firms derive economic power not solely from producing content themselves, but from controlling the infrastructure through which content circulates. This position enables platforms to capture value generated by billions of users whose participation makes the platform economically valuable. In this respect, platform ownership resembles other historical forms of concentrated ownership in which control over productive infrastructure permits extraction of economic rents from the activities of others.\footnote{See generally Brett Frischmann, \textit{Infrastructure: The Social Value of Shared Resources} (2012).}

Grassroots architectures reduce this intermediary position by relocating ownership and possession to participants themselves. Economic relationships need not pass through a central platform because participants retain both the assets and the infrastructure necessary for exchange.

This does not eliminate intermediaries entirely. Applications, search services, reputation systems, discovery mechanisms, and other forms of coordination may continue to emerge. The distinction is that such intermediaries would compete for access to participant-owned assets rather than exercising effective ownership over those assets themselves. Economic power would therefore arise from providing services rather than controlling participation. The result is the possibility of alternative market structures in which value capture occurs closer to the individuals and communities generating value in the first place.

The significance of intermediary rent extraction therefore extends beyond questions of market structure or business models. As Part II argued, platforms function as juridical infrastructures that allocate ownership, possession, and authority through the interaction of copyright, contract, and technical architecture. The rents extracted by platforms are not merely economic outcomes but legal and institutional consequences of those underlying arrangements. Because users lack possession of the infrastructure, surrender broad contractual rights, and participate through systems owned by others, value generated through their expressive activity becomes available for intermediary appropriation.

Viewed in this light, the economic question is inseparable from the institutional one. Where ownership, possession, and governance are concentrated within intermediaries, the value generated through participation becomes available for intermediary capture. Where those elements remain distributed among participants, the legal and economic conditions of participation change accordingly. The question is therefore not simply who earns revenue from digital participation, but which juridical infrastructure governs the production and distribution of that revenue in the first place.

\subsubsection{Limitations and Open Questions}

These economic possibilities remain subject to substantial practical limitations.

Transaction costs remain. Direct licensing and compensation require payment infrastructure, contracting mechanisms, and systems for discovery and enforcement.\footnote{\label{fn:bonneau}See Joseph Bonneau et al., \textit{SoK: Research Perspectives and Challenges for Bitcoin and Cryptocurrencies}, IEEE Symp. on Security \& Privacy 104 (2015).}

Discovery presents additional challenges. Platforms currently provide recommendation systems, search functionality, and audience aggregation. Grassroots architectures must develop alternative mechanisms for connecting creators with audiences without recreating the concentration of power they seek to avoid.

Collective-action and free-rider problems also persist. Open sharing may reduce incentives for payment-based systems, while restrictive licensing may reduce the benefits of network participation. Determining appropriate balances between openness and compensation remains an unresolved design question.

Finally, network effects continue to favor established platforms.\footnote{\label{fn:lemley-network}See Mark A. Lemley \& David McGowan, \textit{Legal Implications of Network Economic Effects}, 86 Cal. L. Rev. 479 (1998).} Ownership alone does not guarantee adoption. Users frequently value convenience, familiarity, and existing social connections. Whether distributed ownership can overcome these advantages remains an empirical question.

Copyright ownership of digital speech acts does not guarantee any particular economic outcome. It does, however, establish the legal preconditions necessary for participants to capture value generated through their own digital activity rather than ceding that value by default to intermediary institutions.
\section{Conclusion}\label{sec:conclusion}

Copyright law is largely neutral regarding architecture. The same legal system that facilitates platform dominance can also support person ownership. Whether ownership remains distributed among participants or becomes concentrated within intermediary institutions depends upon the technical and institutional arrangements through which copyright operates.

Server-based architectures concentrate possession, control, and governance within intermediary institutions. Device-based architectures distribute possession and permit ownership to remain with participants. Copyright law follows those architectural facts. When people create expressions, possess the underlying data, and retain legal rights in their works, ownership remains distributed. When platforms aggregate possession, govern access, and acquire extensive rights through contract, ownership and control become effectively centralized.

The stakes extend beyond social networking. Contemporary platforms have demonstrated that digital participation generates assets capable of producing substantial economic, social, and political value. Questions concerning the ownership of those assets therefore increasingly bear upon questions of individual autonomy, democratic participation, economic agency, and self-government.

Conventional regulatory responses generally accept platforms as inevitable and seek to constrain their power through oversight and accountability mechanisms. Those reforms may be valuable, but they leave largely intact the underlying architecture through which ownership and control are concentrated. Grassroots architectures approach the problem from a different direction: not by regulating concentrated ownership, but by altering how ownership is distributed in the first place. By combining distributed possession, person-owned copyright, and voluntary contractual governance, they create conditions under which digital participation need not depend upon intermediary permission.

This Article's ultimate claim is one of possibility rather than inevitability. Contemporary platforms are not the only architecture through which digital social life can be organized. Existing copyright doctrine is capable of recognizing digital speech acts as copyrightable works, and technical architectures exist in which ownership, possession, and control remain distributed among participants rather than concentrated in intermediaries. Although this Article focuses on digital speech acts within grassroots architectures, the doctrinal analysis developed here may also bear on other forms of cryptographically authenticated digital expression. Whether such systems ultimately prove desirable, effective, or sustainable remains an empirical question.                  


\end{document}